\documentclass[twocolumn]{article}

\usepackage{arxiv}
\usepackage{xurl}
\usepackage[colorlinks=true,allcolors=blue]{hyperref}
\usepackage[utf8]{inputenc} 
\usepackage[T1]{fontenc}    
\usepackage{url}    
\usepackage{tabularx}
\usepackage{array}
\usepackage{booktabs}       
\usepackage{amsfonts}       
\usepackage{nicefrac}       
\usepackage{microtype}      
\usepackage{doi}
\usepackage{makecell}
\usepackage{amsmath}
\usepackage[numbers]{natbib}
\usepackage[capitalise,noabbrev]{cleveref}
\usepackage{graphicx}
\usepackage{cuted}

\usepackage{natbib}
\graphicspath{ {./images/} }
\usepackage{xcolor}
\usepackage[acronym]{glossaries}
\usepackage{fancyhdr} 
\glsdisablehyper  
\usepackage{gensymb}
\usepackage{ulem}
\usepackage{subcaption}
\usepackage{hyperref}

\newcommand{\affit}[1]{$^{\mathrm{\textnormal{\textit{#1}}}}$}

\newacronym{ecmwf}{ECMWF}{European Centre for Medium-Range Weather Forecasts}
\newacronym{ddms}{DDMs}{Data-driven models}
\newacronym{ddm}{DDM}{data-driven model}
\newacronym{eps}{EPS}{ensemble prediction system}
\newacronym{epss}{EPSs}{ensemble prediction systems}
\newacronym{metno}{MET Norway}{Norwegian Meteorological Institute}
\newacronym{metcoop}{MetCoOp}{Meteorological Co-operation on Operational \acrshort{nwp}}
\newacronym{nwp}{NWP}{numerical weather prediction}
\newacronym{synop}{SYNOP}{surface synoptic observations}
\newacronym{nmhses}{NMHSes}{national meteorological and hydrological services}
\newacronym{gnns}{GNNs}{graph neural networks}
\newacronym{gnn}{GNN}{graph neural network}
\newacronym{gpu}{GPU}{graphical processing unit}
\newacronym{gpus}{GPUs}{graphical processing units}
\newacronym{lam}{LAM}{limited area model}
\newacronym{lams}{LAMs}{limited area models}

\newacronym{t}{T}{2-meter temperature}
\newacronym{ws}{WS}{10-meter wind speed}
\newacronym{p6h}{P6h}{6-hour precipitation accumulation}
\newacronym{mslp}{MSLP}{mean sea level pressure}
\newacronym{tmin}{Tmin}{daily minimum 2-meter temperature}
\newacronym{tmax}{Tmax}{daily maximum 2-meter temperature}
\newacronym{wsmax}{WSmax}{daily maximum wind speed}
\newacronym{p24h}{P24h}{24-hour precipitation accumulation}

\newacronym{ifs}{IFS}{Integrated Forecast System}
\newacronym{meps}{MEPS}{MetCoOp Ensemble Prediction System}
\newacronym{aifs}{AIFS}{Artificial Intelligence Forecasting System}

\newacronym{fft}{FFT}{fast Fourier transform}
\newacronym{relu}{ReLU}{rectified linear unit}
\newacronym{mlp}{MLP}{multilayer perceptron}
\newacronym{nlp}{NLP}{Natural Language Processing}

\newacronym{crps}{CRPS}{Continuous Ranked Probability Score}
\newacronym{fcrps}{fCRPS}{Fair Continuous Ranked Probability Score}
\newacronym{ssr}{SSR}{spread-skill ratio}
\newacronym{mse}{MSE}{mean squared error}
\newacronym{mae}{MAE}{mean absolute error}
\newacronym{rmse}{RMSE}{root mean square error}
\newacronym{ets}{ETS}{equitable threat score}
\newacronym{fss}{FSS}{fractions skill score}
\newacronym{dct}{DCT}{discrete cosine transform}
\newacronym{bss}{BSS}{Brier skill score}
\newacronym{qq}{QQ}{quantile-quantile}

\title{HourGlass: A Probabilistic Data-Driven Temporal Downscaler for Hourly Global and Regional Weather Forecasting}

\author{
 Magnus Sikora Ingstad \affit{a*} \\
  \And
 Mariana C. A. Clare \affit{b*} \\
  \And
  Olav Ersland \affit{a} \\
  \And
 Vera Gahlen \affit{b} \\
 \And
 Håvard Homleid Haugen \affit{a} \\
 \And
 Oph\'elia Miralles \affit{a} \\
  \And
 Even M. Nordhagen \affit{a} \\
  \And
 Thomas N. Nipen \affit{a} \\
  \And
 Ivar A. Seierstad \affit{a} \\
  \And
 John Bjørnar Bremnes \affit{a}
  \And
  Michael Maier-Gerber \affit{b} \\  
  \And
  Zied Ben Bouall\`egue \affit{b} \\
\And
Harrison Cook \affit{b} \\ 
\And
Christian Lessig \affit{b} \\
  \And
  Gert Mertes \affit{b} \\
  \And
Cathal O'Brien \affit{b} \\
\And
Florian Pinault \affit{b} \\
\And
Ana Prieto Nemesio \affit{b} \\
\And
 Matthew Chantry \affit{b}
}

\begin{document}

\twocolumn[
\begin{@twocolumnfalse}
\maketitle
\begin{abstract}
Many forecast applications require high-frequency temporal resolution, while most state-of-the-art data-driven weather forecasting systems operate at 6-hourly temporal resolution. Direct hourly forecasting is possible, but leads to error accumulation and temporal consistency issues. Here, we introduce HourGlass, a probabilistic data-driven temporal downscaling methodology for reconstructing the temporal evolution between two forecast states. HourGlass models are trained probabilistically using variants of the continuous ranked probability score (CRPS) to preserve small-scale spatial variability and encourage temporal consistency. This is in contrast to existing temporal downscaling methodologies, which have been trained deterministically and suffer from smoothing. Additionally, by training on forecast trajectories, we overcome temporal inconsistencies in existing reanalysis/analysis datasets that previous approaches were trained on.
In this work, we apply HourGlass in two settings: AIFS-HourGlass (applied in a global setting to AIFS-Single and AIFS-ENS, ECMWF's global data-driven weather forecasting model) and Bris-HourGlass (applied in a high-resolution regional setting to Bris, MET Norway's ensemble stretched-grid data-driven weather forecasting model). Verification against observations shows that both HourGlass models preserve the skill of the underlying forecasting systems while producing temporally coherent hourly forecasts with realistic small-scale variability. The case studies further demonstrate that the HourGlass models can reproduce physically consistent temporal evolution during rapidly evolving weather events, including extratropical cyclones and organised convection. Hourly precipitation remains a challenge, with HourGlass improving the spatial realism of precipitation fields, but continuing to underestimate the most intense precipitation extremes, as is common for data-driven weather forecasting models. These results demonstrate the potential of HourGlass models to bridge the gap between coarse-temporal-resolution data-driven forecasts and the hourly products required in operational regional and global forecasting.

\end{abstract}

\bigskip
\vspace{0.5cm}
\end{@twocolumnfalse}
]

\footnotetext[1]{Norwegian Meteorological Institute, Oslo, Norway}
\footnotetext[2]{ECMWF, Reading, UK}
\footnotetext{%
*Equal Contribution\\
Corresponding authors:\\
\href{mailto:magnusi@met.no}{\texttt{magnusi@met.no}};
\href{mailto:mariana.clare@ecmwf.int}{\texttt{mariana.clare@ecmwf.int}}
}

\section{Introduction}\label{sec:intro}
Recent years have seen a rapid rise in data-driven weather forecasting, with interest from both public meteorological services and the private sector \citep{agrawal_machine_2019, bi_pangu-weather_2022, keisler2022, pathak_fourcastnet_2022,lam_graphcast_2023, oskarsson_graph-based_2023, lang_aifs_2024, price_gencast_2024}. 
Most data-driven weather forecasting models are trained on the European Centre for
Medium-Range Weather Forecasts' (ECMWF) ERA5 reanalysis dataset \citep{hersbach2020era5} and produce forecasts at a 6-hourly temporal resolution. However, for downstream applications and extreme event forecasts, a higher temporal resolution is typically necessary for good forecasts. 

Two main strategies can be considered to bridge this gap: i) training a forecasting model that directly predicts hourly outputs \citep{bi_pangu-weather_2022}, or ii) training a dedicated model that reconstructs intermediate hourly states from two forecast bounding times \citep{leinonen2024modulated, zhong_fuxi-20_2024}. Direct hourly forecasting can be affected by unphysical temporal correlations in the training data and lead to excessive error accumulation \citep{zhong_fuxi-20_2024, leinonen2024modulated}. This makes hourly downscaling from, e.g., 6-hourly machine-learning forecasts a more attractive option. Such a downscaling approach was used in \citet{leinonen2024modulated} (building on the AFNO architecture introduced in FourCastNet \citep{pathak_fourcastnet_2022}) and in \citet{zhong_fuxi-20_2024} (building on the FuXi architecture introduced in \cite{chen_fuxi_2023}). Both models are trained on ERA5 reanalysis data at 0.25° resolution using a latitude-weighted mean squared error objective, and show signs of both spatial smoothing and jumpiness in the forecasts themselves.

In this work, we present HourGlass, a probabilistic data-driven temporal downscaling methodology which we apply in two settings: AIFS-HourGlass (applied in a global setting to AIFS-Single and AIFS-ENS, ECMWF's global data-driven weather forecasting single and probabilistic models \citep{moldovan_update_2025,lang2026aifscrps}) and the stretched-grid Bris-HourGlass (applied in a high-resolution regional setting to Bris, MET Norway's stretched-grid probabilistic model\citep{nordhagen2025}). All models are implemented in the Anemoi Framework \citep{anemoi}, an open-source collaborative framework for developing data-driven weather models.

The HourGlass temporal downscaling methodology introduces two main novelties. Firstly, we train our models probabilistically. Many studies have shown that training a data-driven weather forecasting model towards a mean-squared error (MSE) results in spatially smoothed forecasts \citep{ben2024rise, lam_graphcast_2023,nipen2025}. In the context of a temporal downscaler, this is particularly undesirable because it would lead to temporally inconsistent spatial smoothing, since more smoothing will occur at the centre of the downscaling window where the predictive uncertainty is highest. This smoothing would be particularly prominent when applying a deterministic temporal downscaler to a sharp probabilistic forecaster, as every 6th hour would have much sharper fields. The issue can be avoided by probabilistic training that yields physically more realistic samples \citep{price_gencast_2024, nordhagen2025, lang2026aifscrps}. Therefore, in this work, we train the HourGlass models probabilistically using noise injection and training towards a CRPS loss function, using a similar set-up to \cite{lang2026aifscrps} and \cite{nordhagen2025}. This enables us to produce forecasts with realistic small-scale variability across space and time. Furthermore, we would like the forecasts to be temporally consistent, i.e. the forecast should evolve coherently within each temporal downscaling window without physically unrealistic hour-to-hour fluctuations. We achieve this by augmenting the CRPS loss with additional terms that penalise deviations in the maximum, minimum, and mean values over the window.

The second main novelty of the present work is the dataset we use. Although ERA5 reanalysis data exists at hourly resolution, the 4D-Var method used to create the reanalysis dataset means that it is not temporally consistent across the diurnal cycle. Moreover, most regional reanalysis/analysis datasets, including MET Norway's MEPS analysis dataset, do not exist at an hourly resolution. Hence, in this work, we propose training on forecast data from Numerical Weather Prediction (NWP) models, thereby emulating the temporal evolution of the NWP forecast with our temporal downscaler. As we are applying the temporal downscaler to a data-driven forecaster, we still have the skill of the data-driven forecaster and do not limit ourselves to the skill of the NWP model as we would be if we trained the forecaster directly on forecast data.

The paper is structured as follows. Section~\ref{sec:metho} presents the methodology and datasets, Section~\ref{sec:results} presents verification scores against observations, Section~\ref{sec:case_study} shows case studies of weather events. Finally, Section~\ref{sec:conclusion} summarises the main conclusions and discusses remaining open questions.

\section{Methodology}
\label{sec:metho}
\begin{figure*}
    \centering
    \includegraphics[width=.9\linewidth]{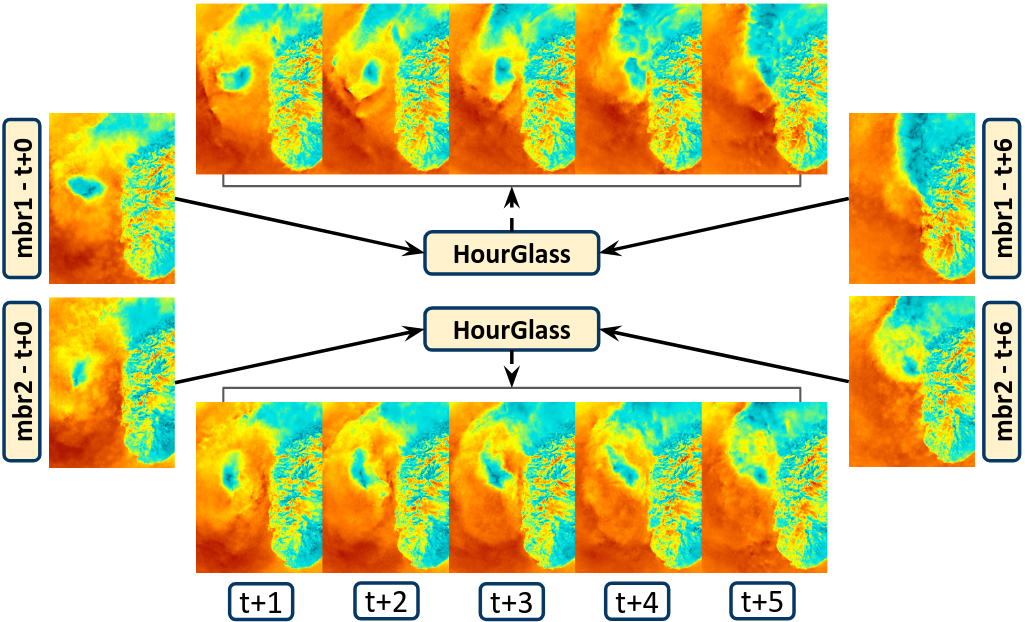}
    \caption{Example of how a forecasting and temporal downscaling model can be used in inference. Pictured at t and t+6 is 10m wind speed for two members of the Bris ensemble forecasting model trained in \citep{nordhagen2025}. Bris-HourGlass is run on the 6-hour interval for each member separately to increase its temporal resolution.}
    \label{fig:forecaster-downscaler_setup}
\end{figure*}

The purpose of the temporal downscaler is to reconstruct the temporal evolution between two forecasting states provided by a forecasting model, rather than to forecast future weather states from a single initial time. Specifically, our model takes as input the states at hours $t_{0}$ and $t_{6}$ from a forecaster and reconstructs the intermediate hourly states (see Figure \ref{fig:forecaster-downscaler_setup}). Some variables, such as precipitation, are diagnostic in the temporal downscaler (i.e., these fields are not provided as inputs by the upstream forecaster to the temporal downscaler but instead diagnosed from the other atmospheric variables). Hence, we also predict the state at $t_{6}$. 

 We consider both the global AIFS-HourGlass at 31 km resolution and the regional Bris-HourGlass at 2.5 km resolution, with the regional model obtained by fine-tuning a global model on a stretched grid, following \citet{nipen2025}. The HourGlass models are probabilistic, with stochasticity introduced through noise injection in the latent space and training towards a probabilistic score, following the AIFS-CRPS approach of \citet{lang2026aifscrps}.

\subsection{Model architecture}
\label{sec:archi}
The temporal downscaler model adopts the same base model architecture as existing 6-hourly forecasting models \citep{lang_aifs_2024, nipen2025, nordhagen2025, lang2026aifscrps,moldovan_update_2025}. The model is implemented within the Anemoi framework \citep{anemoi} and is based on graph neural networks (GNNs), with a graph transformer as the encoder, processor and decoder. In the processor, the stretched-grid Bris-HourGlass has a level 7 graph refinement globally (~50 km) and level 10 refinement regionally (~7 km), following \citep{nordhagen2025}. The global AIFS-HourGlass has a level 6 graph refinement (~100 km). Note that the use of a graph transformer for the processor is different from previous AIFS models which used a transformer \cite{moldovan_update_2025}. Following \citet{lang_aifs_2024, nipen2025, nordhagen2025}, both temporal downscaling models use 1,024 channels and 16 layers in the processor. The main architectural differences to these forecasting models arise from the introduction of a time output dimension. In the decoder, all six output times are produced simultaneously, with the processor-injected latent noise shared across all output time steps. A single forward pass then samples a temporally coherent trajectory. Hence, the computational cost of the temporal downscaler remains close to that of a forecasting model. For prognostic variables, we apply a backward skip connection in physical space, so the model predicts a tendency rather than a full field. As a result, the residual target for the final timestep of the temporal downscaling window is zero, such that the model learns not to change the prognostic variables for the last timestep.

\subsection{Training objective}
\label{sec:training}
It is well known that machine learning weather forecasting models trained towards MSE produce smooth outputs (see, e.g. \cite{ben2024rise}). In a temporal downscaler context, this leads to more smoothing in the centre of the temporal window due to a larger uncertainty there. When training probabilistically, this increased uncertainty is instead represented by more spread at the centre of the downscaling window, whilst still preserving small-scale spatial variability. We therefore train our HourGlass model probabilistically with two ensemble members. 
As shown in Figure~\ref{fig:forecaster-downscaler_setup}, during inference, only a single downscaling realisation between t and t+6 is generated for each forecast member (rather than an ensemble). This also applies to the temporal downscaling of single forecasts such as AIFS-Single.

For our base training objective, we follow the almost-fair \acrshort{crps} loss introduced in \cite{lang2026aifscrps}. Per grid point and target time, this can be decomposed into mean absolute error \acrshort{mae} and variability as
\begin{align}\label{eq:afcrps}
    &\mathcal{L}_{p,t}\left(\{x\},y\right)=  \\
    \frac{1}{M}\sum_{i=1}^M|x_i-y| &-\frac{1-\varepsilon}{2M(M-1)}\sum_{i=1}^M\sum_{j=1}^M|x_i-x_j|, \notag
\end{align}
where $\mathcal{L}_{p,t}$ represents the almost-fair \acrshort{crps} function, ${y}$ is the truth, ${x}$ is the model predicted forecast, $M$ is the number of ensemble members and $\epsilon$ is a configurable parameter. Note that if $\epsilon=0$, the above equation is the fair CRPS; here we use $\epsilon = 0.025$ following \cite{lang2026aifscrps}.

To promote temporal consistency (i.e. coherent evolution of the forecast within each temporal downscaling window), we also incorporate statistics of the output timesteps across the time window in the loss. Specifically, we penalise the model for large differences between output timesteps, and for deviations in minimum, maximum and mean across the time window. We do this by computing the almost fair CRPS (\ref{eq:afcrps}) on time-aggregated values of the prediction and of the truth. We compute the additional terms of
\begin{equation}\label{eq:time_aggregate}
\mathcal{L}_{p,t}\left(\{f(\text{prediction})\},f(\text{truth})\right),
\end{equation}
where $f$ is $\min$, $\max$, $\operatorname{mean}$ and $\operatorname{difference}$ defined as
\begin{equation}\label{eq:agg_difference}
    \operatorname{difference}(\textbf{p}) = (p_2 - p_1, ..., p_6-p_5),
\end{equation}
and add these to the almost fair CRPS for our loss function.

\begin{figure}
    \centering
    \includegraphics[width=\linewidth]{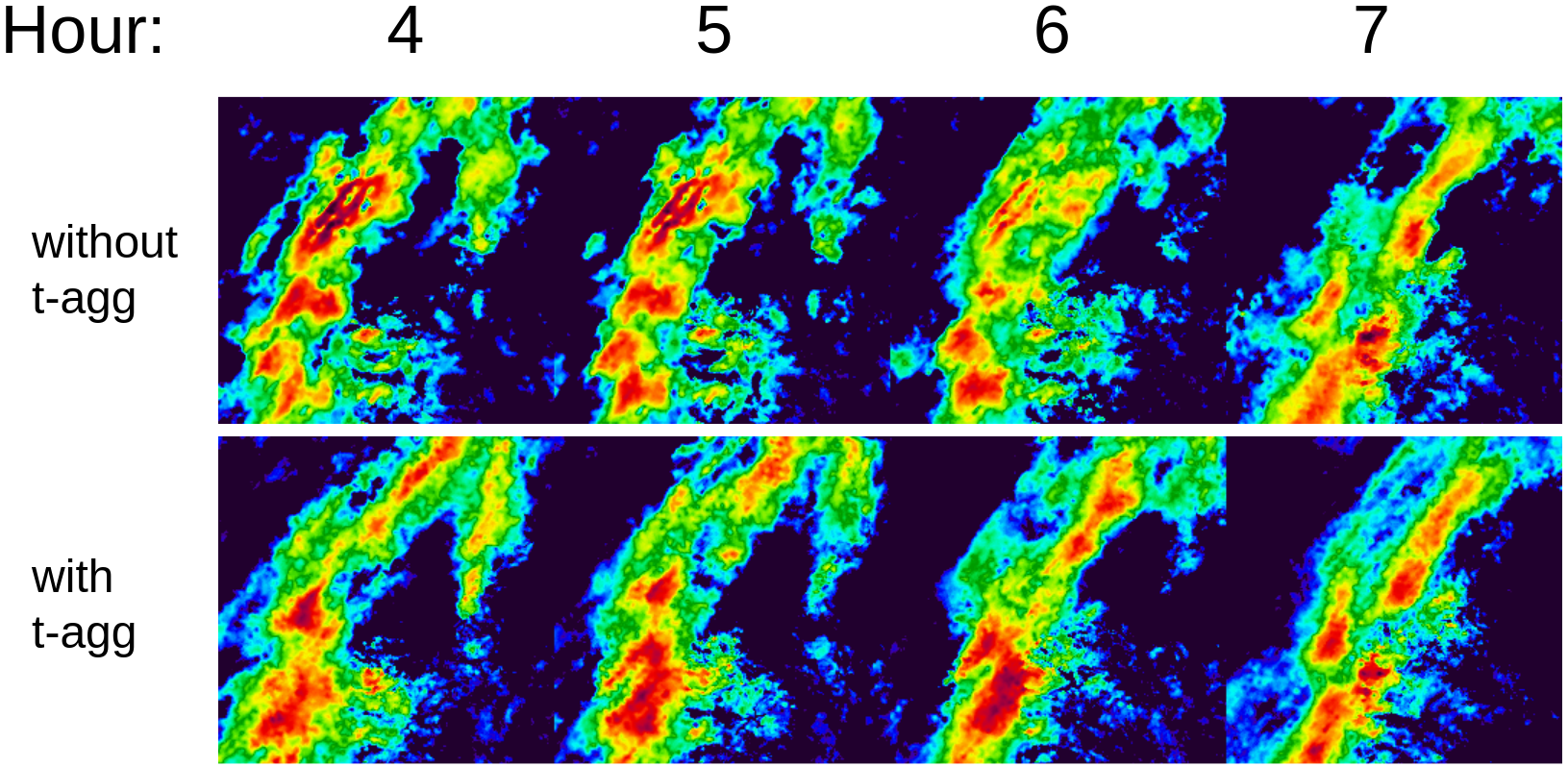}
    \caption{Hourly precipitation forecast off the west coast of Norway produced by Bris-HourGlass trained with and without $\min$, $\max$, $\operatorname{mean}$ and $\operatorname{difference}$  loss terms. Hours 4-6 share the same random noise.}
    \label{fig:time_consistency}
\end{figure}

Figure \ref{fig:time_consistency} illustrates the added value of these terms. When the extra loss terms are not applied, hours 4-6 share similar features that change abruptly at hour 7. Adding the extra terms in the loss enforces a smooth evolution in precipitation patterns within the same 6-hour period. Additionally, with more variability between each intermediate hour, the change with the new initialisation at hour 7 is less abrupt.

For Bris-HourGlass, we include an additional spatial consistency term, computing the two-dimensional \acrfull{fft} of $\{\textbf{x}\}$ and $\textbf{y}$ over the regular-gridded Nordic region. This is known to improve fine-scale spatial structures \citep{nordhagen2025}. For the regional subset of points, we get the term
\begin{equation}\label{eq:fft}
    \mathcal{L_{\text{freq}}} = \sum_{p,t}\mathcal{L}_{p,t}\left(\text{FFT}(\{\textbf{x}\}),\text{FFT}(\textbf{y})\right).
\end{equation}
Additionally, we compute the loss over $\left(\text{FFT}(\{f(\textbf{x}\})),\text{FFT}(f(\textbf{y}))\right)$, with the time-aggregation function $f$ in \eqref{eq:agg_difference}.

\subsection{Datasets}
\label{sec:data}

\begin{figure*}
    \centering
    \includegraphics[width=0.8\linewidth,
        clip]{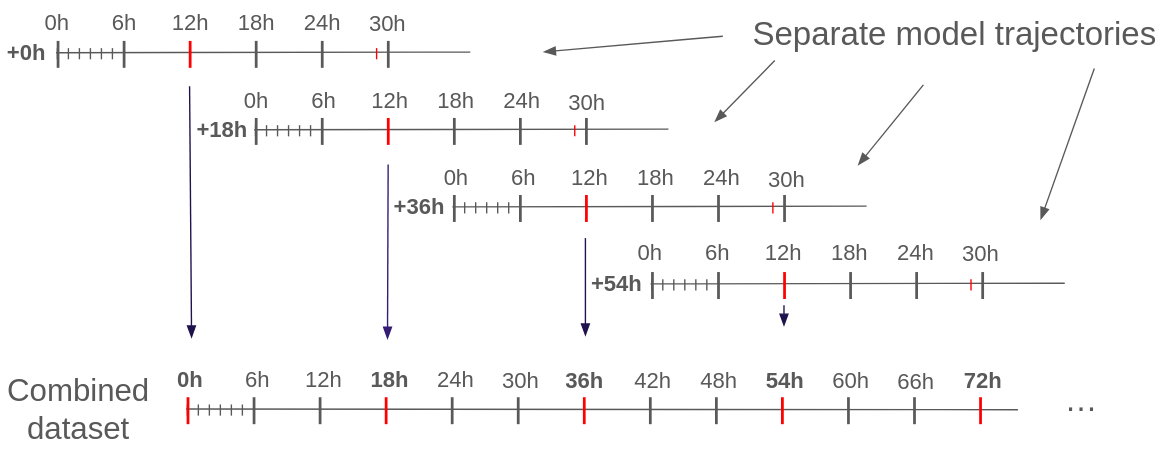}
    \caption{Construction of forecast datasets for training. A training sample must stay within the boundaries marked in red.}
    
    \label{fig:forecast_dataset_creation}
\end{figure*}

\subsubsection{Global Datasets}
In order to train our temporal downscaler effectively, we require target data at an hourly resolution. ECMWF’s ERA5 global reanalysis dataset \citep{hersbach2020era5} is available at hourly temporal resolution on its native N320 grid (approximately 31 km spatial resolution).

\begin{figure*}
    \centering
    \includegraphics[width=0.75\linewidth]{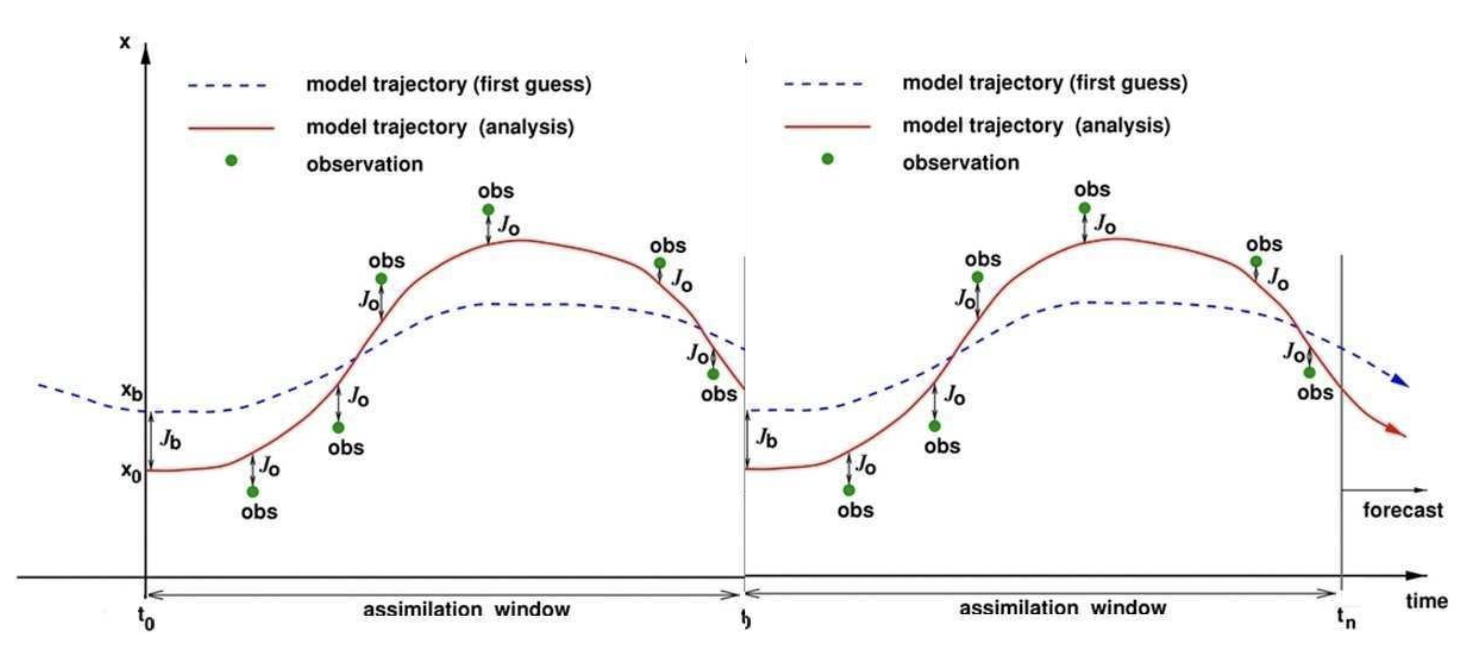}
    \caption{Schematic illustration of jumpiness introduced by 4D-Var assimilation cycling. Stitching forecasts from adjacent 12-hour assimilation cycles can introduce discontinuities at the cycle boundaries, 09 UTC and 21 UTC.}
    \label{fig:era5_jumps}
\end{figure*}

ERA5 has been used as a basis for training most global data-driven weather forecasting models (see for example \cite{lam_graphcast_2023, lang_aifs_2024}). However, a limitation of ERA5 for temporal downscaling is that the reanalysis is produced through successive 4D-Var assimilation cycles, each solving a separate optimisation problem over a 12-hour window (09-21 UTC and 21-09 UTC) with a different set of observations (see Figure \ref{fig:era5_jumps}). As a result, the reanalysis is not perfectly continuous across analysis updates, and discontinuities can appear at the boundaries between consecutive assimilation windows.

When training a temporal downscaler on ERA5, the model learns these artificial jumps (see Figure~\ref{fig:era5_forecast}). To mitigate this issue, here we train on forecast data which is temporally more consistent within a given forecast integration. For our global forecast dataset, we use ECMWF's operational IFS forecasts, which are initialised at 00, 06, 12, and 18 UTC, with 00 and 12 UTC being the main longer forecast cycles and 06 and 18 UTC shorter supplementary runs \citep{ecmwf_ifs_2024-1, ecmwf_ifs_2024-2}.

\subsubsection{Regional Datasets} 
The regional dataset used in this work originates from the MetCoOp ensemble prediction system (MEPS) \citep{eresma_metcoop_2026}. MEPS is a regional convection-permitting ensemble prediction system based on the HARMONIE–AROME framework \citep{termonia_aladin_2018, frogner2019a}, designed for short-range forecasting over the Nordic region \citep{frogner_convectionpermitting_2019}. MEPS analysis is produced through a limited-area data-assimilation system as a 3D-Var-based framework with ensemble perturbations, and are therefore not subject to the jumpiness of ERA5. However, the data does not exist hourly, as the operational system is refreshed every 3h. Forecasts are then run from these analyses on the same 3-hour cycle, providing short-range regional predictions at convection-permitting resolution. For the training dataset, we use the initialisations at 00, 06, 12 and 18 UTC of the MEPS control forecast.

\subsubsection{Training Samples Construction}
During training, we restrict the data to continuous forecast segments to avoid exposing the model to the forecast initialisation jumps themselves. This allows the model to see all times of day under temporally smooth conditions, while preventing it from learning artificial jumps associated with forecast updates.
Forecast datasets such as IFS and MEPS are indexed by two temporal coordinates: the forecast initialisation time and the lead time. Due to technical constraints associated with the version of the \texttt{anemoi-datasets} Python package available at the time \citep{anemoi}, only training data with one temporal coordinate was supported. Thus, the datasets were constructed iteratively by extracting temporally continuous forecast segments separated by their length, as illustrated in Figure~\ref{fig:forecast_dataset_creation}. The length of the forecast segments was chosen as 18 hours to ensure that all combinations of lead times and times of day can be seen in the training data. Furthermore, to avoid the spin-up period in the MEPS forecasts \citep{eresma_metcoop_2026}, we did not start from lead time 0h. Instead, for each forecast reference time, we used lead times 12h--29h from IFS and 6h--23h from MEPS forecasts (which uses +6h IFS as its boundary condition). Repeating this procedure over the full archive yields a single continuous valid-time sequence assembled from forecast segments. The same procedure was applied to both the IFS and MEPS forecast datasets.

\subsection{Training setup}
\begin{table*}
 \caption{List of variables used when training the model. Pressure level variables are at 50 hPa, 100 hPa, 150 hPa, 200 hPa, 250 hPa, 300 hPa, 400 hPa, 500 hPa, 600 hPa*, 700 hPa, 800 hPa, 850 hPa, 925 hPa, and 1,000 hPa levels. Prognostic variables are input and output, diagnostic variables are output only, forcings are input only.}
  \centering
  \begin{tabular}{lll}
    \toprule
    Pressure level variables & Single level variables & Forcings \\
    \midrule
    Geopotential height & Skin temperature & Solar insolation \\
    Temperature & 2 m temperature & Sine of Julian day \\
    Specific humidity & 2 m dew point temperature & Sine of latitude \\
    Wind speed u component & 10 m wind speed u component & Sine of local time \\
    Wind speed v component & 10 m wind speed v component & Sine of longitude \\
    Wind speed w component & Mean sea level pressure & Cosine of Julian day \\
    & Surface air pressure & Cosine of longitude \\
    & Total column integrated water & Cosine of latitude \\
    & 1-hour accumulated total precipitation\textdagger & Cosine of local time \\
    & 1-hour accumulated convective precipitation*\textdagger & Land sea mask \\ 
    & Low, Medium, High, Total cloud coverage\textdaggerdbl & Surface geopotential height \\
    & Surface solar radiation downwards\textdagger  \\                              
    & Surface thermal radiation downwards\textdagger  \\                              
    \bottomrule
    \multicolumn{3}{l}{* only in global model} \\
    \multicolumn{3}{l}{\textdagger\, diagnostic variable} \\
    \multicolumn{3}{l}{\textdaggerdbl\, diagnostic variable in global and prognostic variable in stretched grid model} \\
    
  \end{tabular}
  \label{tab:vars}
\end{table*}

\paragraph{AIFS-HourGlass} 
The global AIFS-HourGlass is trained using two ensemble members. Each ensemble member runs on a single GPU and distributed data parallelism is used to enable a global batch size of 8. In total, AIFS-HourGlass is trained over 16 Nvidia GH200 GPUs each with 96GB of memory for a wall time of approximately 120 hours. It is trained to full convergence, using a Cosine Learning Rate scheduler with an effective learning rate of 8e-4, for 200,000 iterations, and with the Adam Optimizer. The model is trained to output 111 meteorological variables, outlined in Table \ref{tab:vars}. To balance the contribution of variables with different temporal variability, we normalise the contribution of variables in the loss function according to the variability of their temporal changes. The loss contribution of each variable is weighted by the inverse of the standard deviation $\sigma$ of its tendencies,
\[
w_i = \frac{1}{\sigma(\Delta x_i)},
\]
where $\Delta x_i(t) = x_i(t+\tau) - x_i(t)$
denotes the tendency of variable $i$ over the forecast interval $\tau$ and the statistics are computed over the training periods.
This is approximately equivalent to optimising errors in tendency space rather than state space. We also apply a pressure level scaler, following standard practice for data-driven weather models (see for example \cite{moldovan_update_2025}).

\begin{figure}
    \centering
    \begin{subfigure}[b]{0.4\textwidth}
        \centering
        \includegraphics[width=\textwidth]{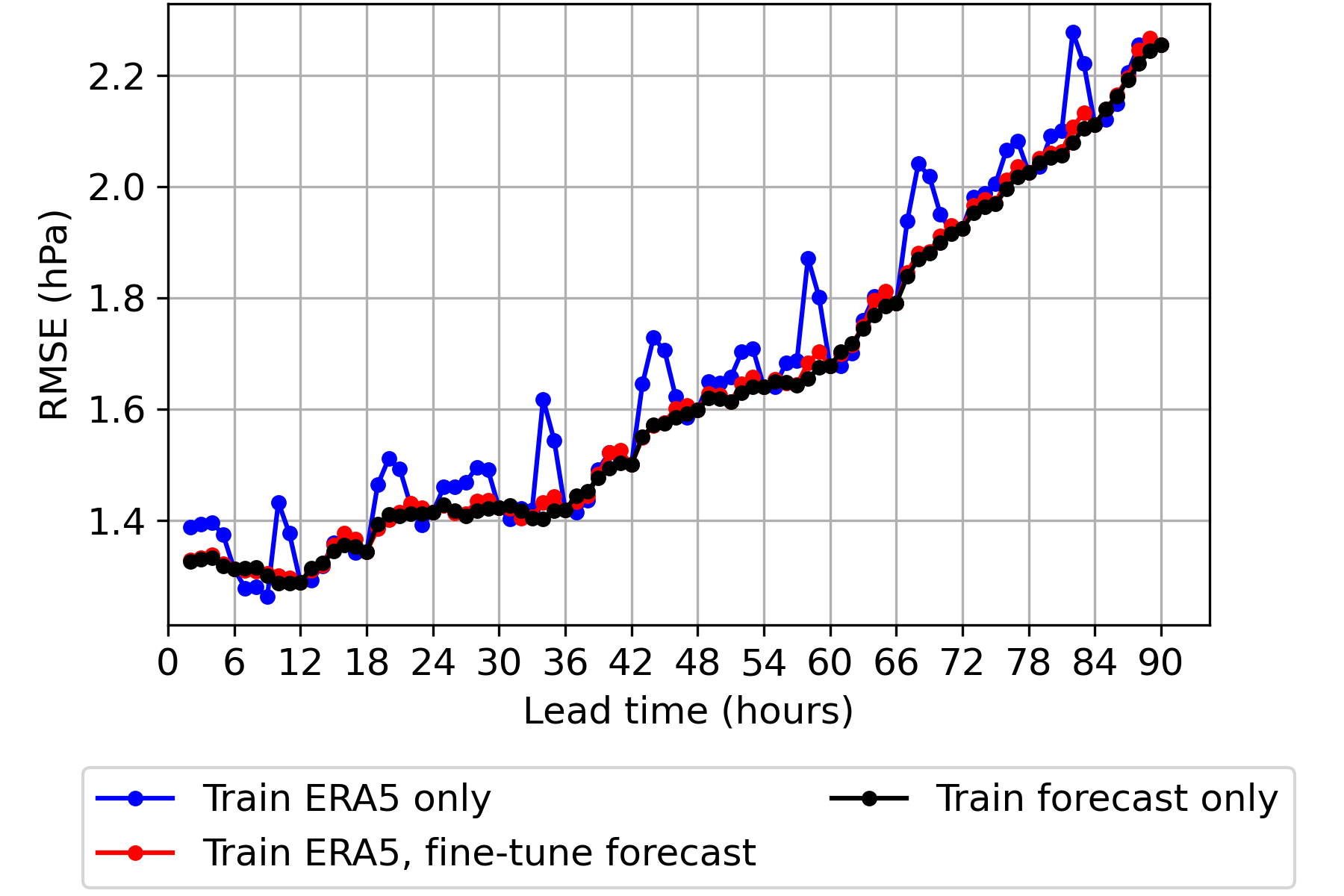}
        \caption{Mean sea-level pressure (northern hemisphere extratropics).}
    \end{subfigure}
    \hfill
    \begin{subfigure}[b]{0.4\textwidth}
        \centering
        \includegraphics[width=\textwidth]{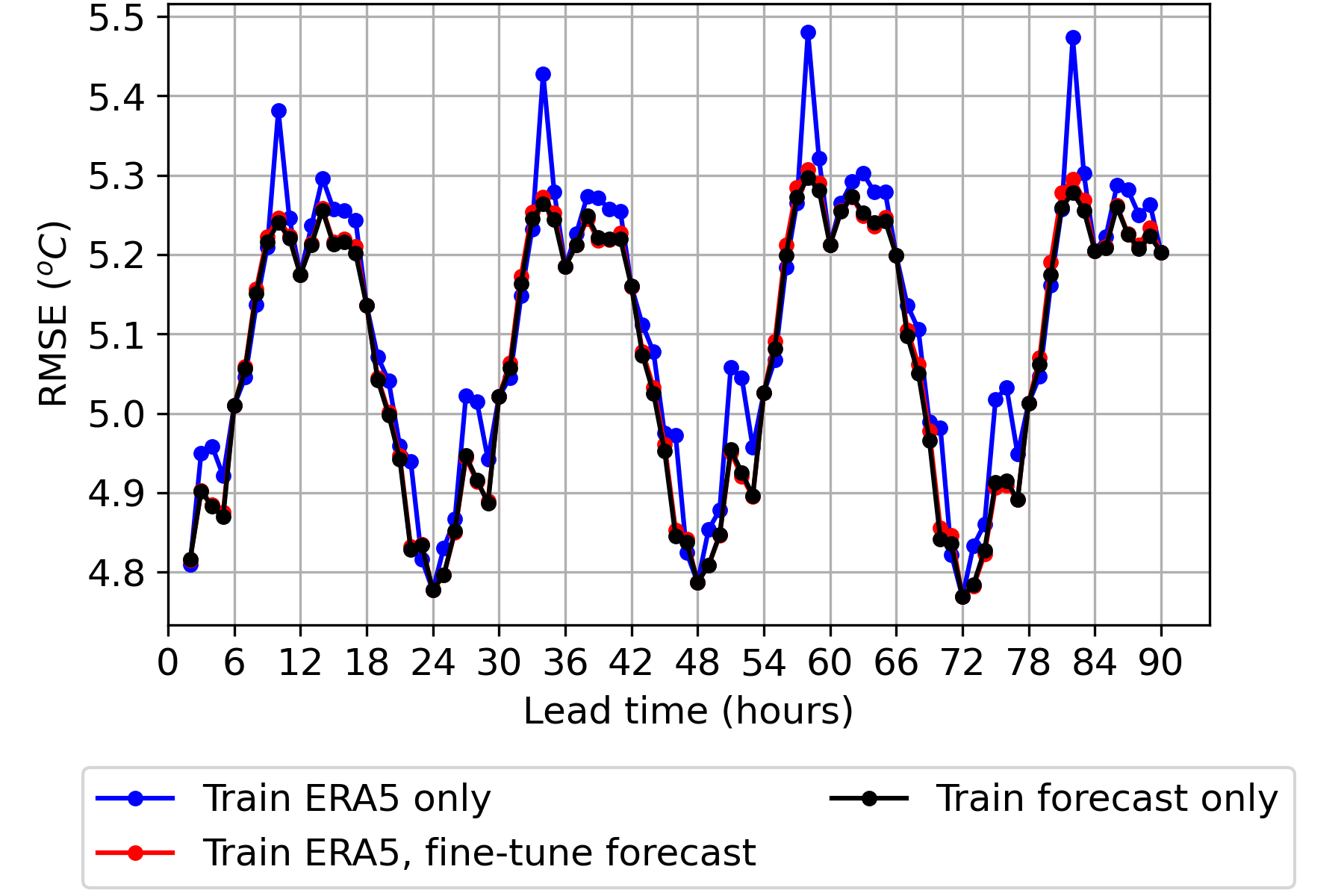}
        \caption{2m temperature (tropics).}
    \end{subfigure}
    \hfill
    \begin{subfigure}[b]{0.4\textwidth}
        \centering
        \includegraphics[width=\textwidth]{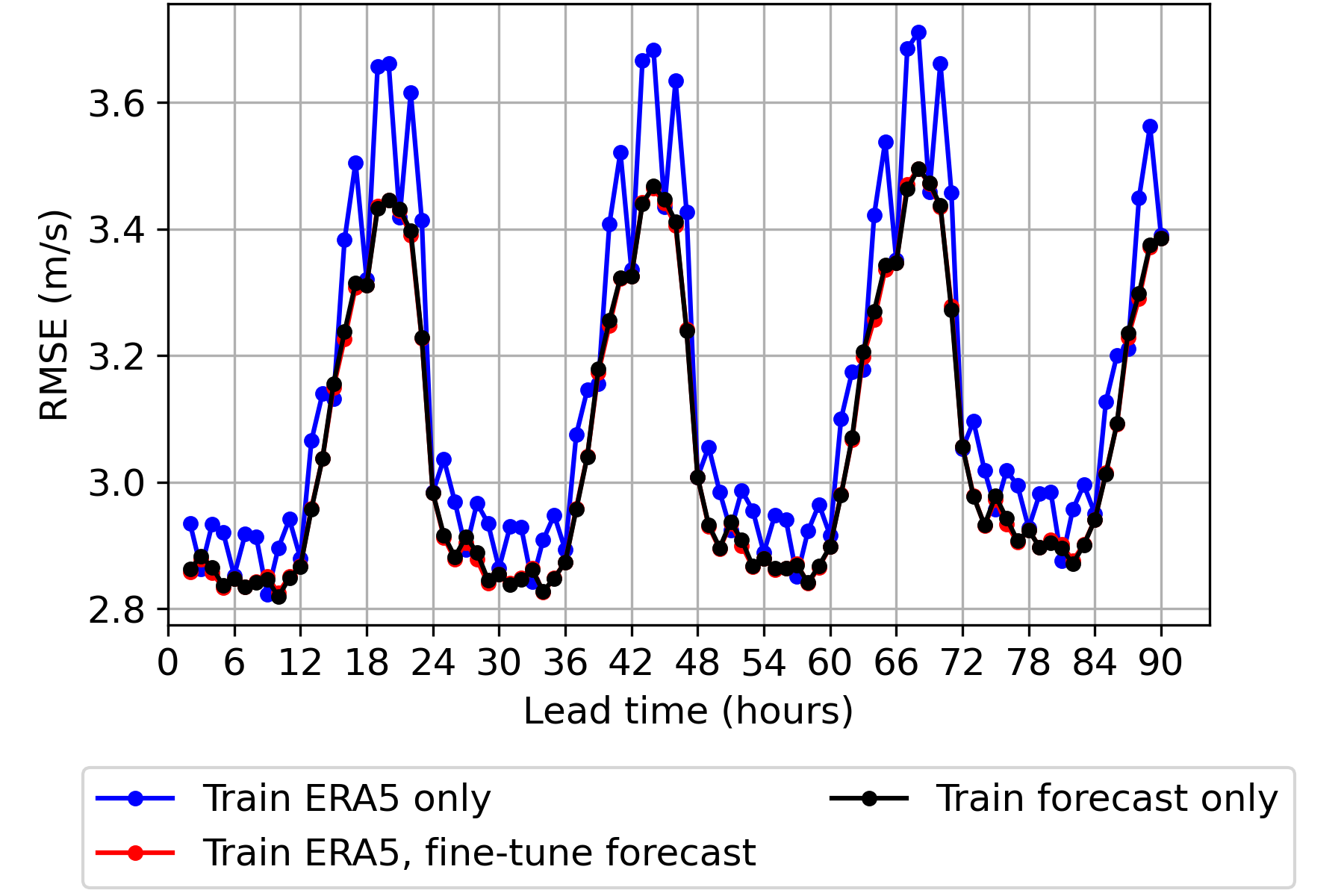}
        \caption{10m wind speed (southern hemisphere extratropics).}
    \end{subfigure}
    \caption{
    RMSE of hourly forecasts produced by AIFS-HourGlass applied to AIFS Single. Three different training approaches are shown: (i) train on ERA5 only; (ii) pre-train on ERA5 and fine-tune on IFS forecast data; (iii) train only on IFS forecast data.
    }
    \label{fig:era5_forecast}
\end{figure}

We explored two different approaches to train with forecast data: (i) pre-train on ERA5 and fine-tune on IFS forecast data; (ii) train only on IFS forecast data. Note that the IFS forecast data has been regridded from the native resolution of TCO1279 (approximately 9 km) to N320 resolution (approximately 31 km). Figure~\ref{fig:era5_forecast} shows that the resulting scores of methods (i) and (ii) are very similar in temporal consistency and we therefore adopt the simpler setup and train the temporal downscaling models using IFS forecast data only. The training period spans 1 January 2016 to 31 May 2022, covering IFS cycles 41r1 to 47r3. Validation is performed over 1 June 2022 to 31 May 2023, corresponding to IFS cycle 47r3.

\paragraph{Bris-HourGlass}
The regional Bris-HourGlass follows the high-resolution stretched grid setup of \citet{nordhagen2025}. To save compute and access more years of training data, we pre-train Bris-HourGlass on global forecast data following the multi-stage approach shown to be effective in \citet{nipen2025}. Fine-tuning Bris-HourGlass using the global AIFS-HourGlass model already presented is, unfortunately, not possible as MEPS does not have the 600 hPa pressure level variables. We follow however the same learning rate scheduler and optimiser as AIFS-HourGlass, and in the pretraining phases train 200,000 iterations on IFS forecast data regridded to O96 resolution (approximately 100 km), training two ensemble members with an effective learning rate of 3.2e-3 and a batch size of 16. We then fine-tune on 50,000 iterations of IFS forecast data regridded to N320 resolution (approximately 31 km), resetting the scheduler with a reduced effective learning rate of 6.4e-4. For this global pre-training we use the years 2016-2023 for training, and 2024 for validation. Lastly, we reset and run with the same scheduler for 50,000 iterations on the stretched grid of N320 IFS and 2.5 km MEPS. This stage uses 6 February 2020 to 31 December 2023 as training and 2024 for validation. During training, the global and regional domains are weighted equally in terms of point-wise CRPS in \eqref{eq:afcrps}, both including the time-dependent loss in \eqref{eq:time_aggregate}. Additionally, the CRPS on the spectral transform of the regional domain in \eqref{eq:fft} is weighted with $\lambda_f = 0.15$ relative to the total point-wise CRPS.

Bris-HourGlass is trained on NVIDIA A100 GPU nodes, each with 64 GB of GPU memory. For the 100 km resolution global pre-training, ensemble-member sharding is used and each GPU contains 1 ensemble member; training takes approximately 48 hours across 8 nodes. For the 31 km resolution global pre-training, we shard each member across 1 full node, training for approximately 80 hours on 32 nodes. The final stretched grid stage is also sharded across 1 full node, with further chunking of the encoder/processor/decoder to fit within memory. This final stage takes approximately 173 hours across 32 nodes. As this is the most expensive part of training, we note that training for 1/3 the number of iterations on the final stage yields almost equal performance.

\section{Verification Scores} 
\label{sec:results}
Throughout this section, we verify the hourly outputs from our HourGlass models against a subset of surface observations from SYNOP stations available at ECMWF and MET Norway, for the year 2025. Verification against analysis is not possible due to the lack of hourly data and verification against reanalysis is not trustworthy due to the lack of temporally consistent hourly data (see Section \ref{sec:data}). Recall from Section \ref{sec:archi} that for prognostic variables, the 6-hour forecast is unchanged by the HourGlass models, hence the scores at every 6th hour are not pertinent.

\subsection{AIFS-HourGlass}
First, we consider the global setting and apply the same global AIFS-HourGlass model to both AIFS Single V1 and AIFS ENS V1 (see \cite{moldovan_update_2025} and \cite{lang2026aifscrps} respectively for details of these models). 
\begin{figure}
    \centering
    \begin{subfigure}[b]{0.4\textwidth}
        \centering
        \includegraphics[width=\textwidth]{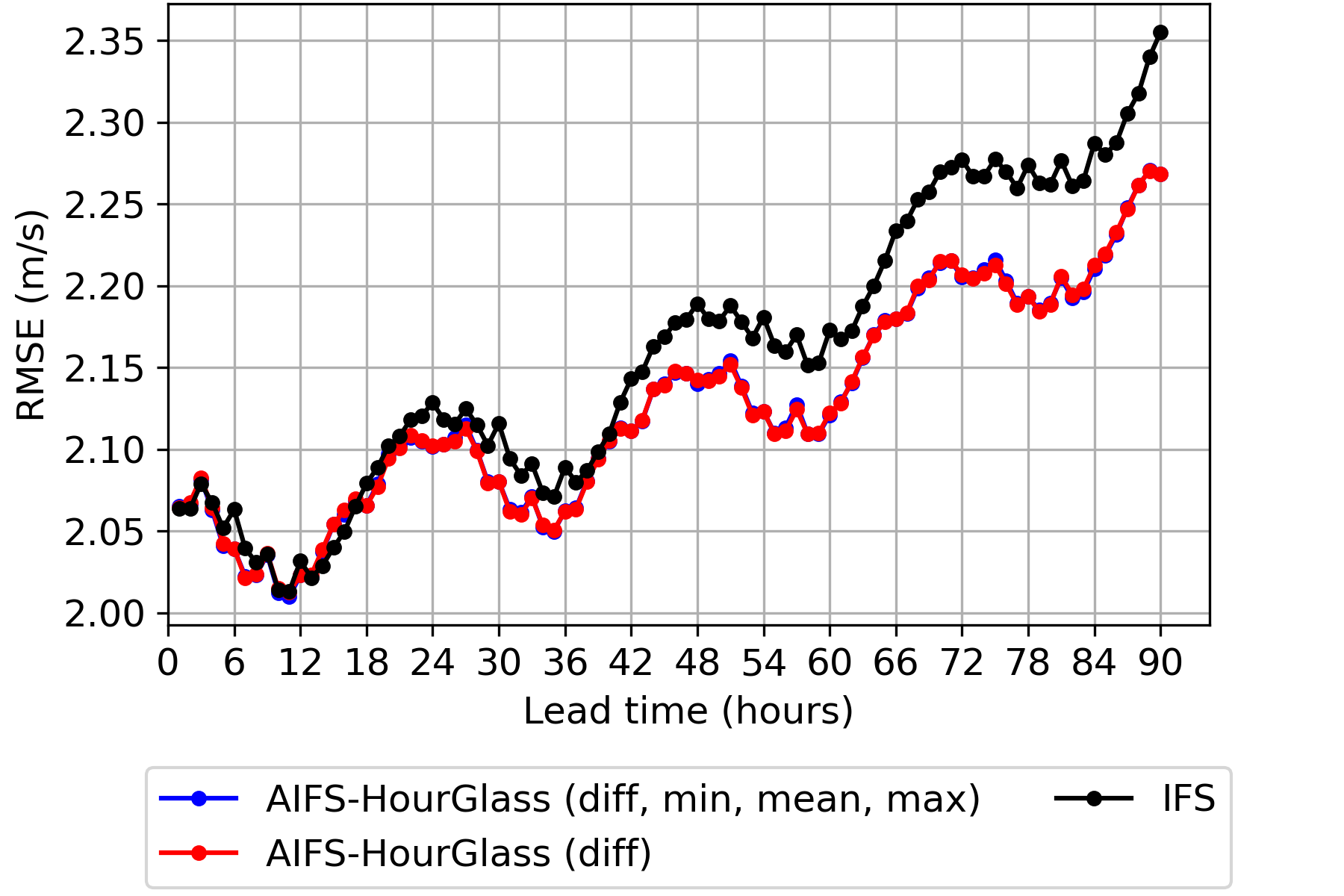}
        \caption{10m wind speed (northern hemisphere extratropics).}
        \label{fig:10uv_single}
    \end{subfigure}
    \hfill
    \begin{subfigure}[b]{0.4\textwidth}
        \centering
        \includegraphics[width=\textwidth]{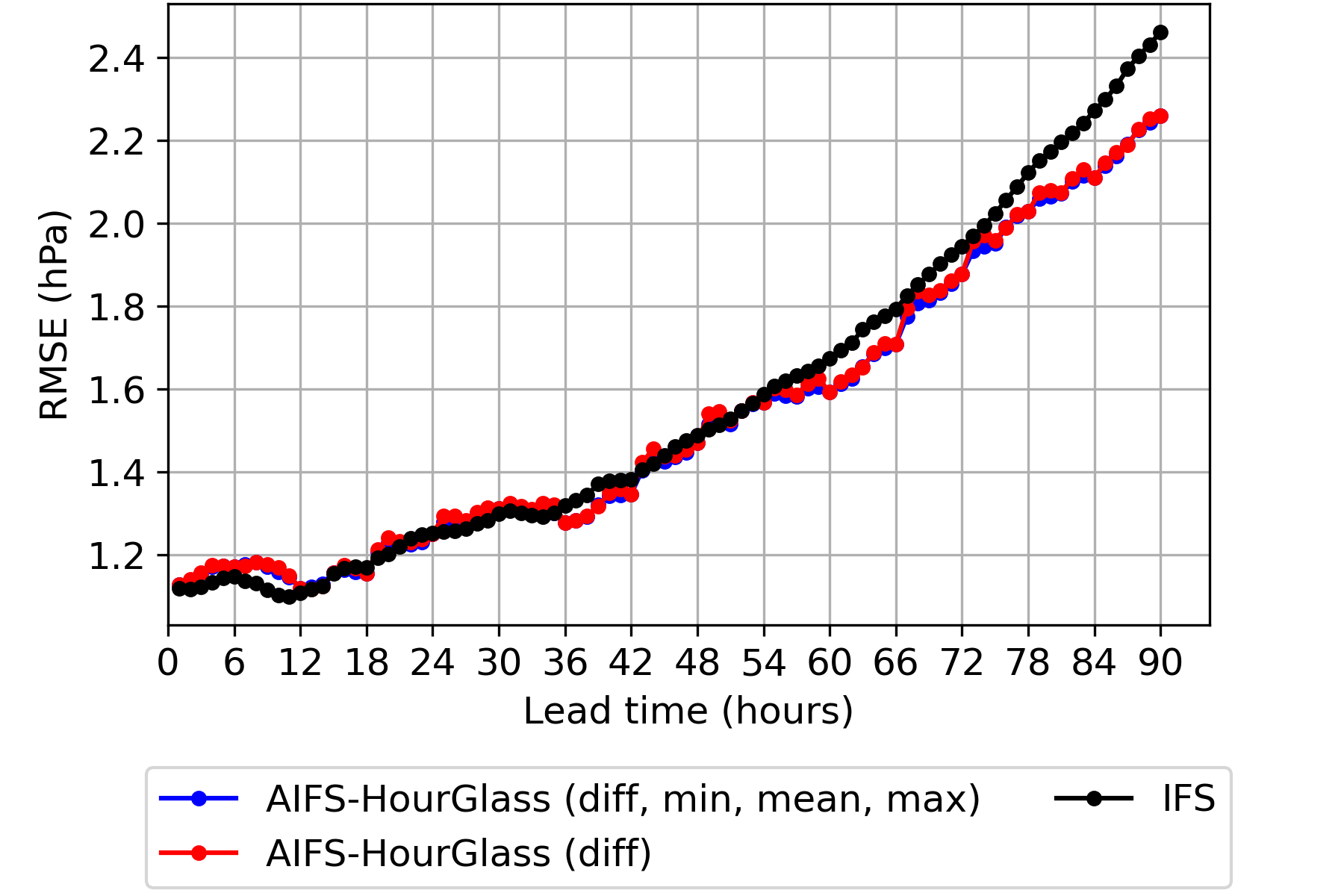}
        \caption{Mean sea-level pressure (northern hemisphere extratropics).}
        \label{fig:rmse_msl_single}
    \end{subfigure}
    \hfill
    \begin{subfigure}[b]{0.4\textwidth}
        \centering
        \includegraphics[width=\textwidth]{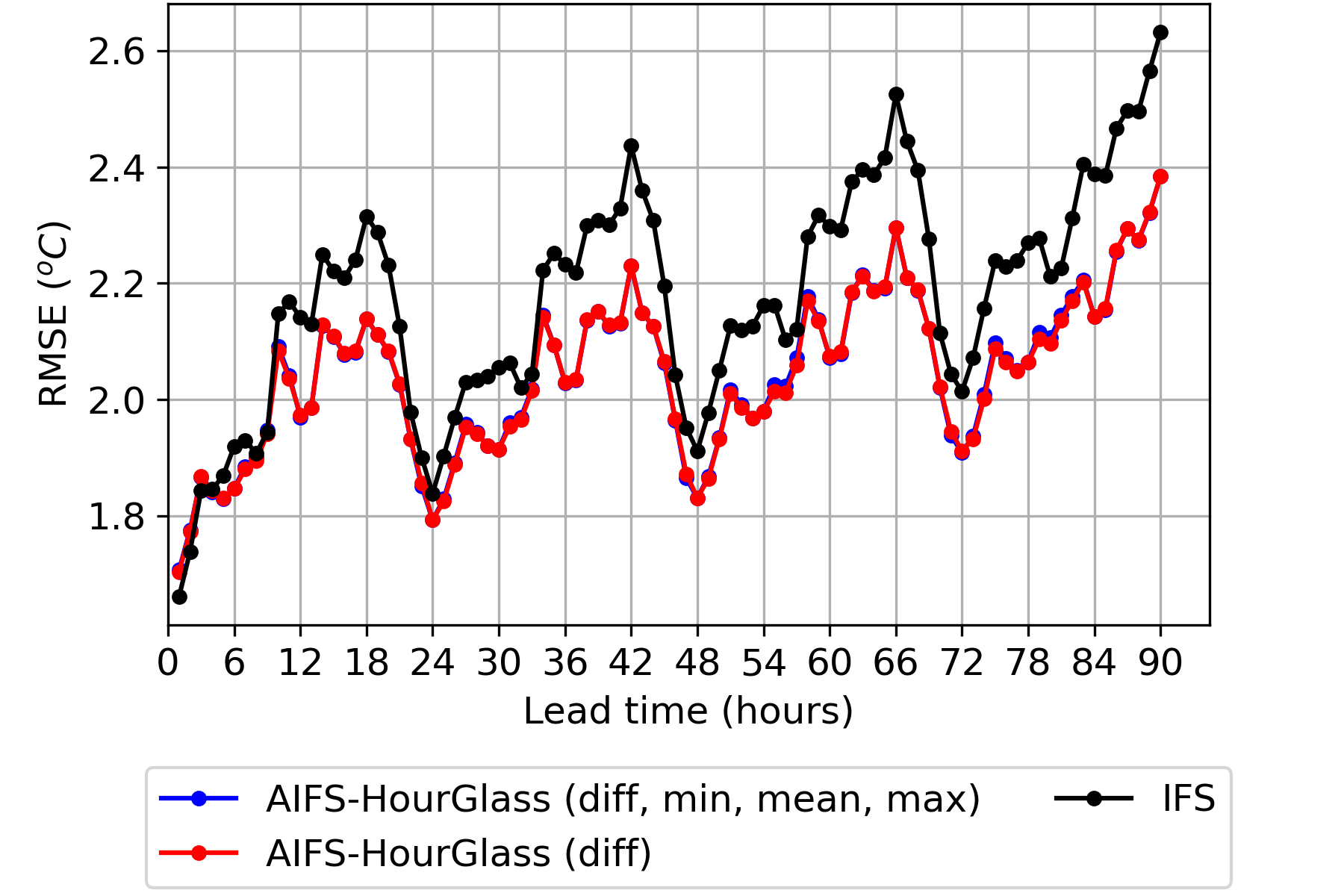}
        \caption{2m temperature (southern hemisphere extratropics).}
        \label{fig:rmse_2t_single}
    \end{subfigure}
    \caption{
    RMSE of hourly forecasts produced by AIFS-HourGlass applied to AIFS Single, compared to IFS hourly forecasts. Here two temporal downscaler models are shown, one where the loss function includes the aggregate $\operatorname{difference}$ term (\ref{eq:agg_difference}) in the total loss and the other where the loss term includes the aggregate $\min$, $\max$ and $\operatorname{mean}$ aggregate terms defined by (\ref{eq:time_aggregate}), as well as the aggregate difference term (\ref{eq:agg_difference}).
    Verification is performed against SYNOP stations for 2025. 
    }
    \label{fig:rmse_global_td}
\end{figure}
\begin{figure}
    \centering
    \begin{subfigure}[b]{0.4\textwidth}
        \centering
        \includegraphics[width=\textwidth]{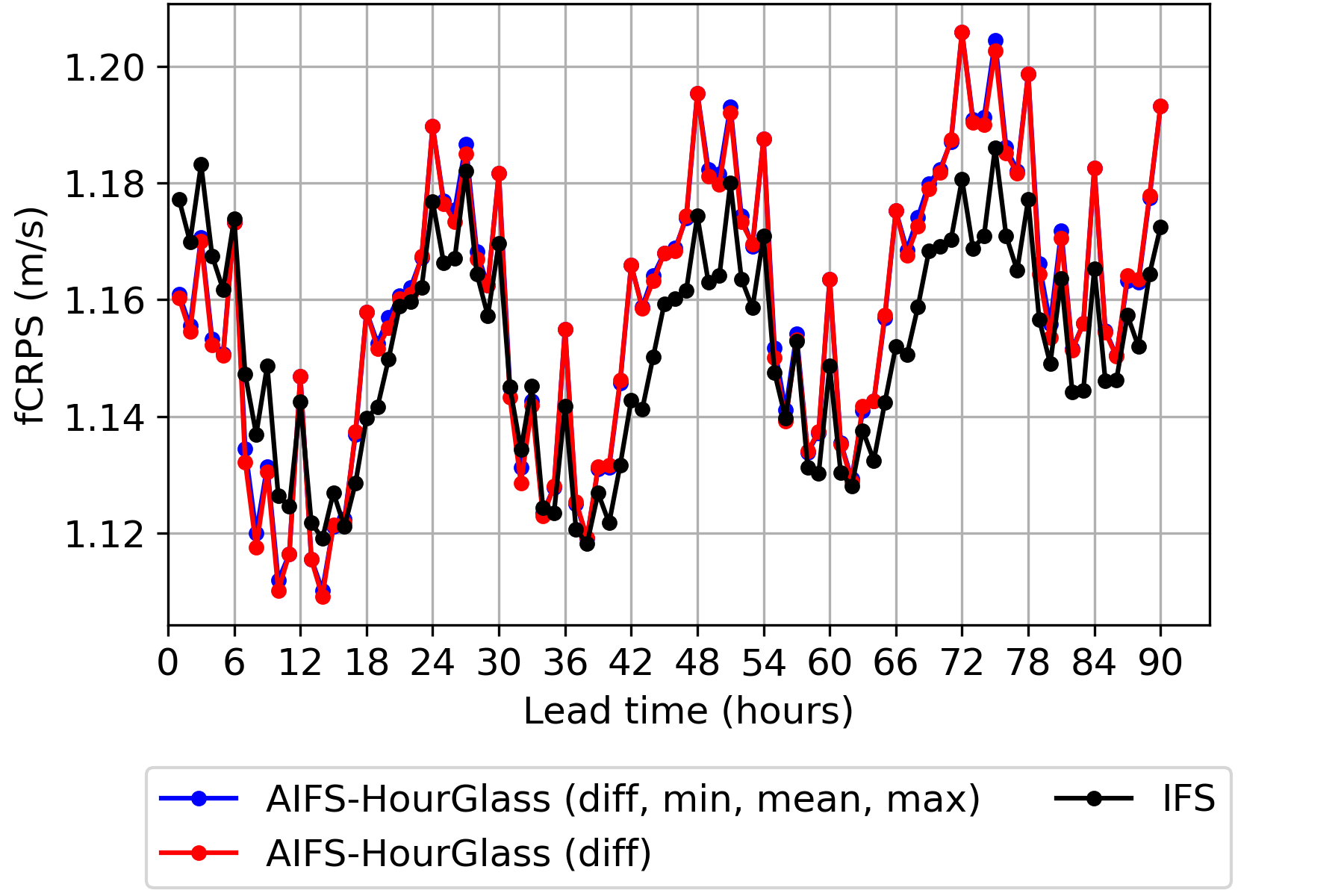}
        \caption{10m wind speed (northern hemisphere extratropics).}
        \label{fig:crps_10uv_single}
    \end{subfigure}
    \hfill
    \begin{subfigure}[b]{0.4\textwidth}
        \centering
        \includegraphics[width=\textwidth]{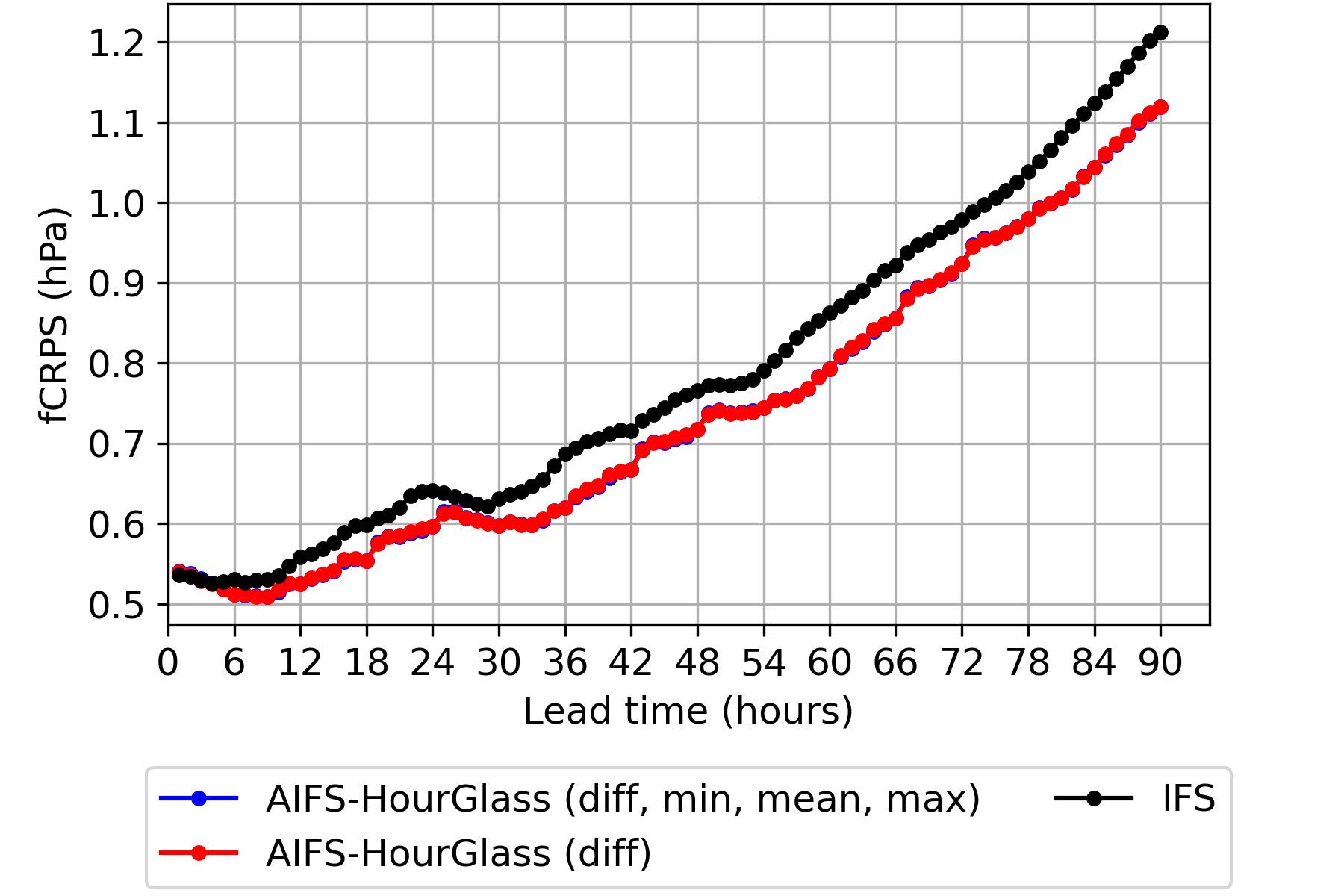}
        \caption{Mean sea-level pressure (northern hemisphere extratropics).}
        \label{fig:crps_msl_single}
    \end{subfigure}
    \hfill
    \begin{subfigure}[b]{0.4\textwidth}
        \centering
        \includegraphics[width=\textwidth]{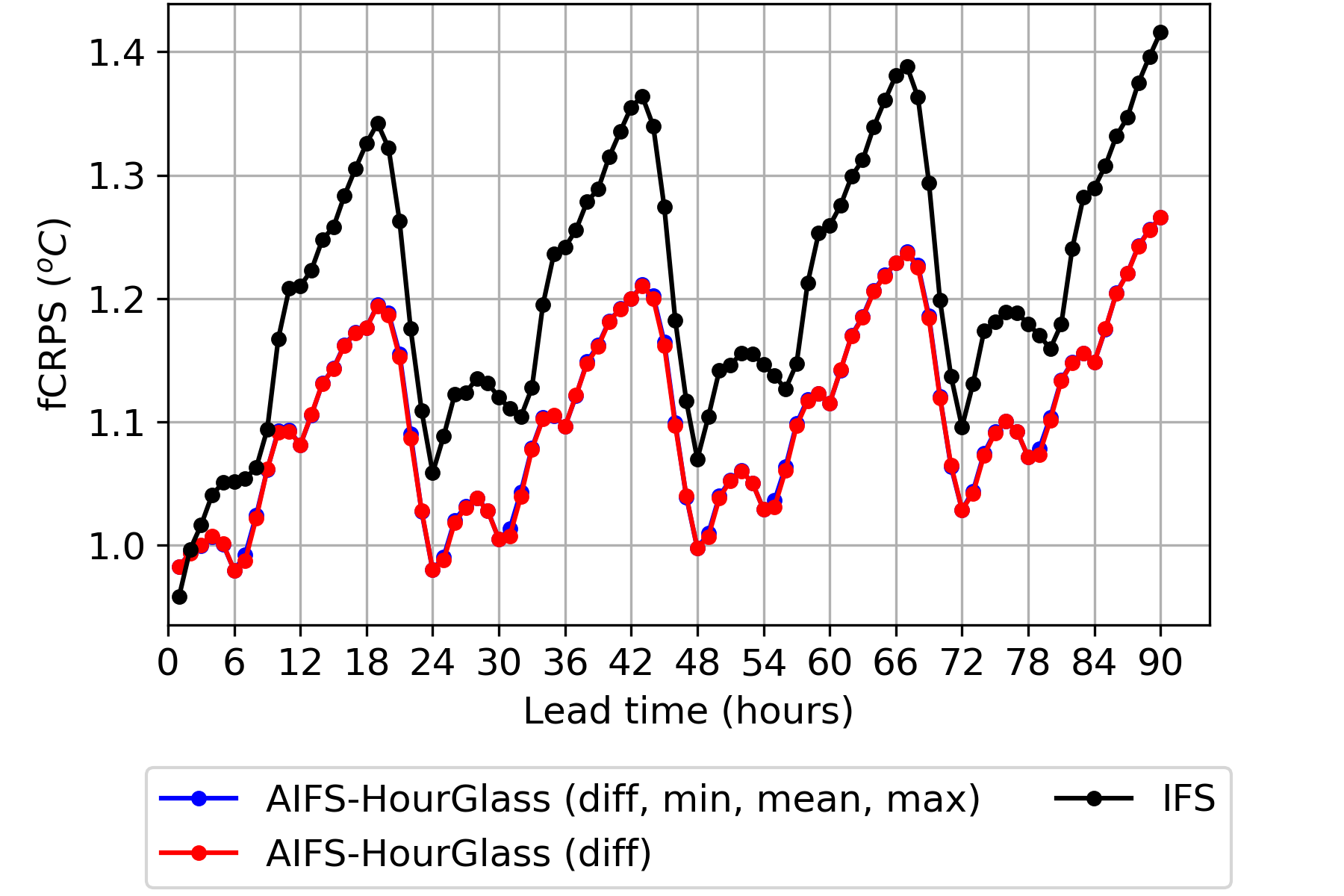}
        \caption{2m temperature (southern hemisphere extratropics).}
        \label{fig:crps_2t_single}
    \end{subfigure}
    \caption{
    Fair CRPS of hourly forecasts produced by AIFS-HourGlass applied to AIFS ENS, compared to IFS hourly forecasts. Here two temporal downscaler models are shown, one where the loss function includes the aggregate $\operatorname{difference}$ term (\ref{eq:agg_difference}) in the total loss and the other where the loss term includes the aggregate $\min$, $\max$ and $\operatorname{mean}$ aggregate terms defined by (\ref{eq:time_aggregate}), as well as the aggregate difference term (\ref{eq:agg_difference}).
    Verification is performed against SYNOP stations for 2025. 
    }
    \label{fig:crps_global_td}
\end{figure}

Figures~\ref{fig:rmse_global_td} and~\ref{fig:crps_global_td} show that applying AIFS-HourGlass to AIFS Single and AIFS-ENS preserves the skill of the underlying forecasting system. The hourly RMSE evolution for AIFS Single and the hourly CRPS evolution for AIFS-ENS closely follow those of the hourly IFS benchmark throughout each 6-hour downscaling window. This temporal consistency indicates that the downscaled states remain consistent with the upstream AIFS 6-hourly forecasts, rather than introducing artificial discontinuities within the window. The small jumps visible in the RMSE evolution of both IFS and temporally downscaled AIFS are largely due to variations in the number of available observations at different times of the diurnal cycle.

\begin{figure*}[h]
    \centering
    \begin{subfigure}[b]{0.48\textwidth}
        \centering
        \includegraphics[width=\textwidth]{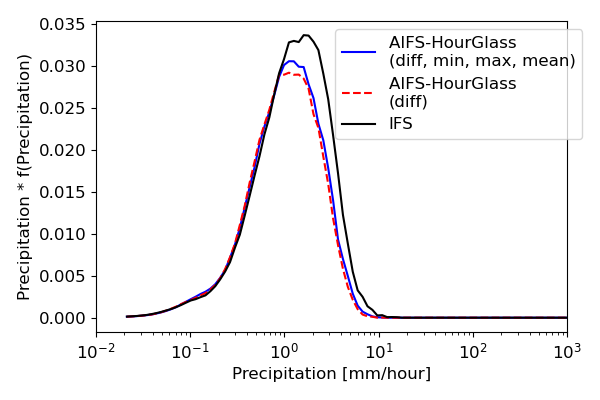}
        \caption{Lead time 5h}
    \end{subfigure}
    \hfill
    \begin{subfigure}[b]{0.48\textwidth}
        \centering
        \includegraphics[width=\textwidth]{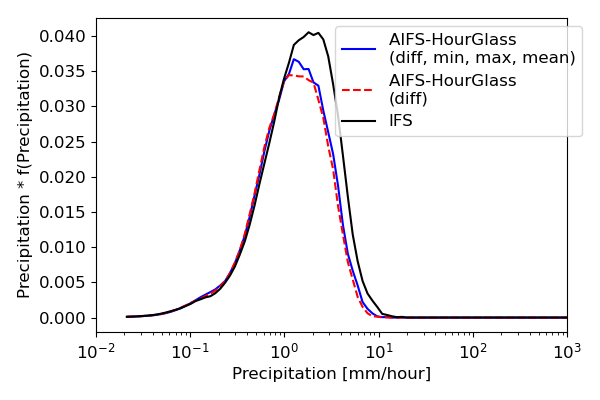}
        \caption{Lead time 6h}
    \end{subfigure}
    \caption{Contribution of each rain rate to the total accumulated precipitation (precipitation * f(precipitation) where f(.) denotes the fraction of precipitation values in each bin) for forecasts in January 2025, comparing IFS and temporally downscaled AIFS Single with different loss functions.  If the curve is shifted right (higher intensities), the model produces more heavy rain; shifted left means more light drizzle.}
    \label{fig:tp_dist}
\end{figure*}

As discussed in Section~\ref{sec:training}, we introduce additional terms to the loss function to encourage temporal consistency. In this section, we analyse the impact of these loss functions on the forecasts. We consider two cases: (i) add only the aggregate $\operatorname{difference}$ term (\ref{eq:agg_difference}) to the total loss; (ii) add the $\min$, $\max$ and $\operatorname{mean}$ aggregate terms defined by (\ref{eq:time_aggregate}), as well as the aggregate difference term (\ref{eq:agg_difference}) to the loss function. Figures \ref{fig:rmse_global_td} and \ref{fig:crps_global_td} show that the choice of temporal consistency loss has little impact on the overall forecast skill for mean sea-level pressure, 2m temperature, and 10m wind speed, with no statistically significant differences observed between the two temporal downscaling configurations for either AIFS Single or AIFS ENS. However, differences become apparent when considering the hourly precipitation distribution (Figure~\ref{fig:tp_dist}). Including the aggregate terms $\min$, $\max$, and $\operatorname{mean}$ defined by (\ref{eq:time_aggregate}) produces a distribution closer to the observed precipitation. Based on these results, we restrict the remaining evaluation of the HourGlass models trained with the $\min$, $\max$, and $\operatorname{mean}$ aggregate terms defined in (\ref{eq:time_aggregate}), together with the aggregate-difference term in (\ref{eq:agg_difference}).

Despite this improvement, both configurations underestimate hourly precipitation relative to IFS (Figure~\ref{fig:tp_dist}). As described in Section~\ref{sec:metho}, precipitation differs from the other fields because it is not provided as an input to the temporal downscaler. It is instead predicted diagnostically, and therefore, the forecast differs between the forecaster and the temporal downscaler. Furthermore, precipitation has a highly non-Gaussian distribution and, in general, lower predictability than other fields.

\begin{figure}
    \centering
    \includegraphics[width=0.48\textwidth]{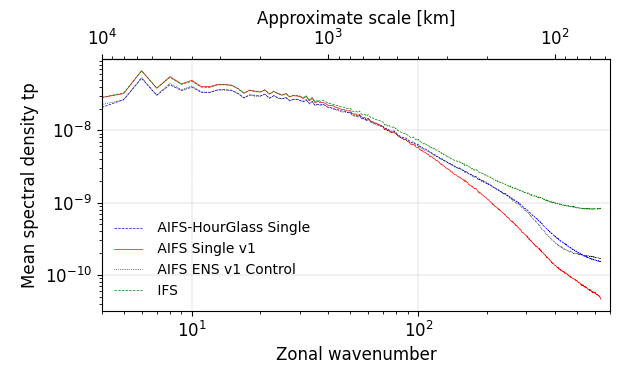}
    \caption{Comparison of spectra of precipitation at a lead time of 6 hours for AIFS-HourGlass Single, AIFS Single v1, control forecast of AIFS ENS v1 and IFS.}
    \label{fig:spectra_tp}
\end{figure}

\begin{figure*}
    \centering
    \includegraphics[width=\linewidth]{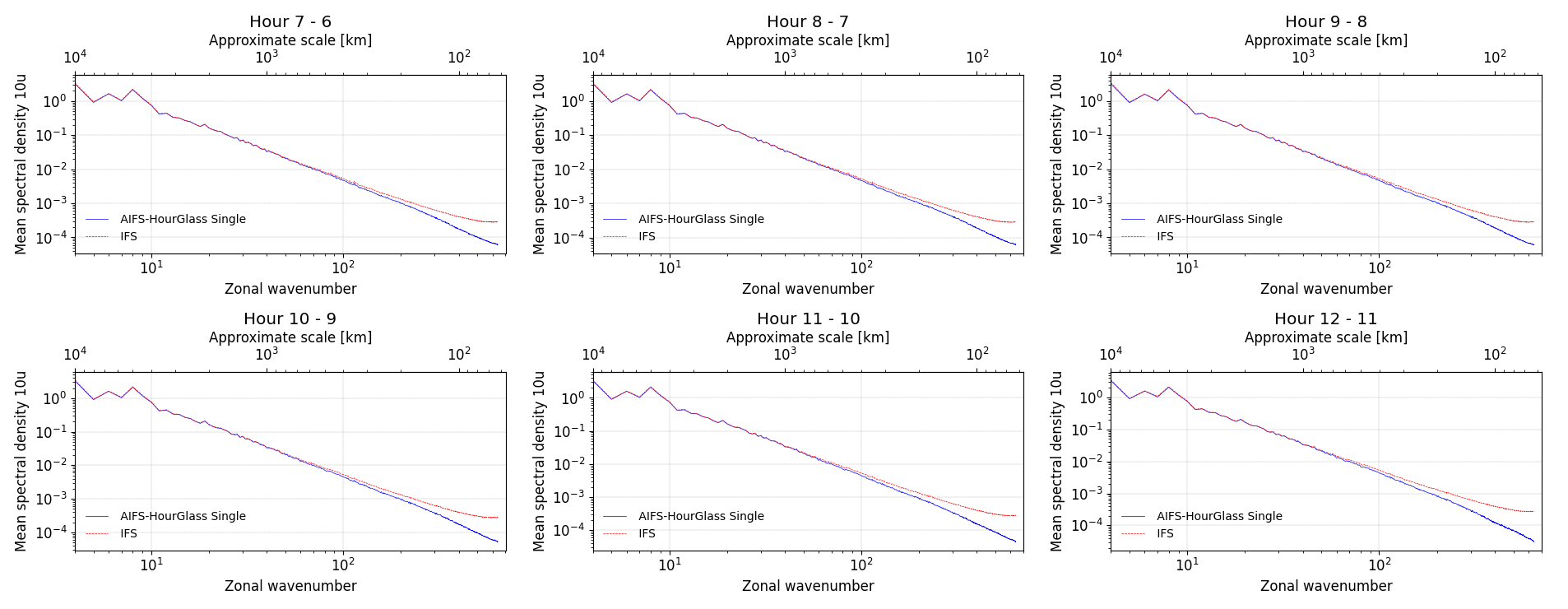}
    \caption{Comparison of spectra of temporal increment of 10m u-wind velocity of AIFS-HourGlass Single and IFS. These show no difference in the spectra of temporal increments across the window, indicating that AIFS-HourGlass is not smoothing the forecast.}
    \label{spectra_surf}
\end{figure*}

In Figure~\ref{fig:spectra_tp}, we compare the spectra of 6-hourly accumulated precipitation from AIFS Single v1, the AIFS-ENS v1 control forecast, IFS, and the temporally downscaled versions of AIFS Single v1 with all loss terms. The AIFS-ENS v1 control forecast is a single realisation of AIFS-ENS initialised from the unperturbed IFS analysis \citep{lang2026aifscrps}, and therefore can be interpreted as a representative member of the ensemble.
The spectra show that the AIFS-HourGlass forecast contains more power at small spatial scales than the AIFS Single forecast from which they are derived. The small-scale variability is closer to that of the AIFS-ENS v1 control forecast, and the resulting precipitation fields are substantially less smooth than those from AIFS Single v1. However, the AIFS-HourGlass forecasts also exhibit reduced power at larger spatial scales (note we only observe this reduction for precipitation). A similar reduction is visible in the AIFS-ENS v1 control forecast, suggesting that this damping may be related to the use of the CRPS loss. A possible mitigation would be to adopt the multiscale CRPS formulation proposed by \citet{lang_multi-scale_2025}; we leave this investigation to future work.

\begin{figure}
    \centering
    \includegraphics[width=\linewidth]{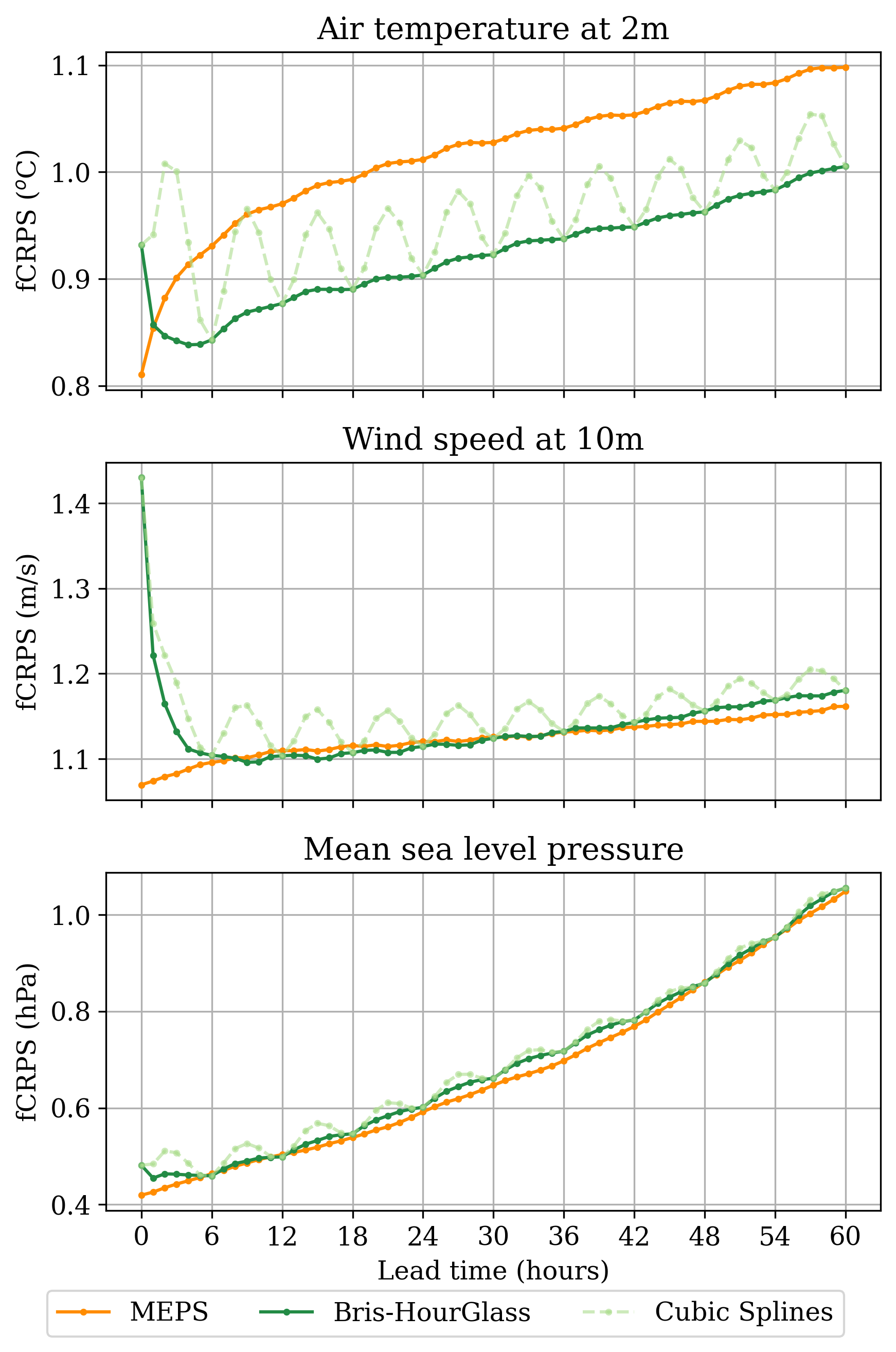}
    \caption{Fair CRPS of hourly forecasts produced by Bris-HourGlass for 2m temperature, 10m wind speed, and mean sea-level pressure. Bris-HourGlass is benchmarked against the MEPS ensemble forecast and a cubic-spline temporal interpolation baseline. Verification is performed against SYNOP for 2025.}
    \label{fig:meps_crps}
\end{figure}

The previous results considered the AIFS-HourGlass forecasts as individual snapshots, which does not directly assess the temporal structure of the reconstructed sequence. We therefore examine the spectra of temporal increments within each downscaling window, in order to evaluate whether the model reproduces realistic hour-to-hour variability. Results for the 10m zonal wind component are shown in Figure~\ref{spectra_surf}.
The sixth step is anchored to the underlying forecaster state through the grid-space residual connection. The consistency of the increment spectra across the full window, including at the transition from the AIFS forecast to the reconstructed sequence, indicates that the step transition is spectrally smooth, with no artificial increase in small-scale power. This further shows that CRPS-based training preserves realistic temporal variability, rather than producing the excessive smoothing one obtains with MSE-trained temporal downscaling methods. This leads to sharper hourly fields and a more physically realistic temporal evolution.

\subsection{Bris-HourGlass}
\label{sec:bris-hourglass}

Next, we consider the regional setting by applying Bris-HourGlass to 10 members of the Bris ensemble forecast trained in \citep{nordhagen2025}. Note that following the ablation study on loss functions for the global model, we train Bris-HourGlass with the $\min$, $\max$, and $\operatorname{mean}$ aggregate terms defined in (\ref{eq:time_aggregate}), together with the aggregate-difference term in (\ref{eq:agg_difference}). We compare Bris-HourGlass to 10 members of the MEPS ensemble forecast, along with a cubic-spline interpolation baseline of the Bris ensemble forecast. 

Figure \ref{fig:meps_crps} shows that the temporal evolution of the forecast from Bris-HourGlass closely follows that of the MEPS ensemble forecast, with only mean sea level pressure scoring slightly worse. The figure also shows that a simple cubic spline interpolation scheme is able to capture the temporal evolution of mean sea level pressure reasonably well, but not that of 2m temperature or 10m wind speed, which both evolve more rapidly in time and contain finer-scale spatial features.

Bris-HourGlass performs particularly well for 2m temperature. It is shown in \cite{nipen2025} that much of the improvement in 2m temperature for the data-driven forecaster is because it is trained on analysis with assimilated temperature observations. Even though Bris-HourGlass is trained on forecast trajectories without such assimilated observations, we see that the intermediate hours also maintain the skill improvement. These results suggest that HourGlass can be used to extend the skill of an analysis to a higher temporal frequency.

\begin{figure}
    \centering
    \includegraphics[width=\linewidth]{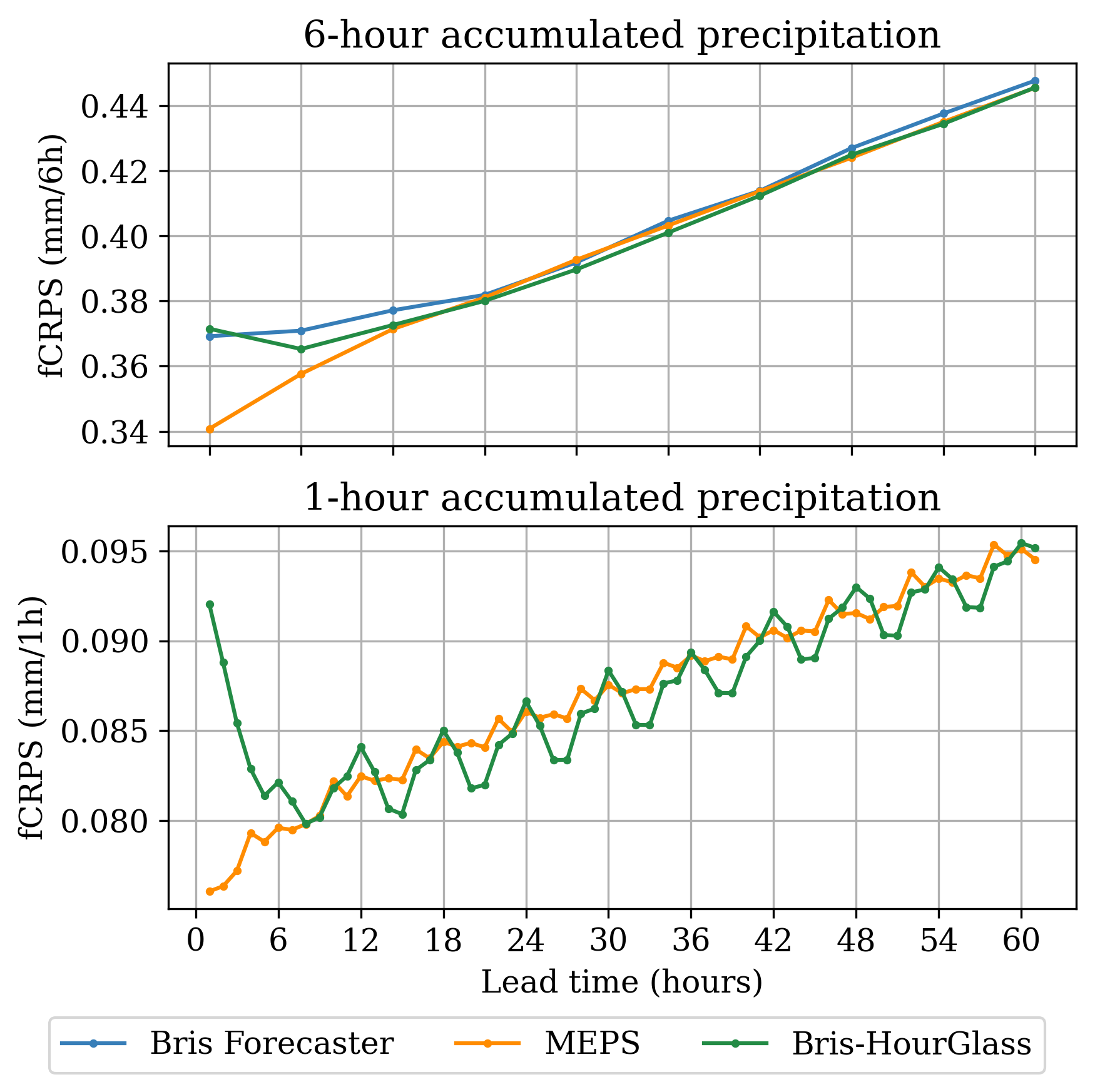}
    \caption{Fair CRPS for hourly and 6-hourly accumulated precipitation for Bris-HourGlass, Bris and the MEPS ensemble forecast. The 6-hourly precipitation is aggregated from the hourly precipitation output from Bris-HourGlass, and compared to the 6-hourly precipitation produced by its upstream forecasting model. Verification is performed against SYNOP for 2025.}
    \label{fig:precip_crps}
\end{figure}

For precipitation, the results are more complex. Figure \ref{fig:precip_crps} shows the fair CRPS for both hourly accumulated precipitation output natively by Bris-HourGlass and 6-hourly accumulated precipitation. Comparing the 6-hourly accumulated precipitation with that from the Bris ensemble forecast, we see that Bris-HourGlass improves upon the scores from the forecasting model.

\begin{figure}
    \centering
    \includegraphics[width=\linewidth]{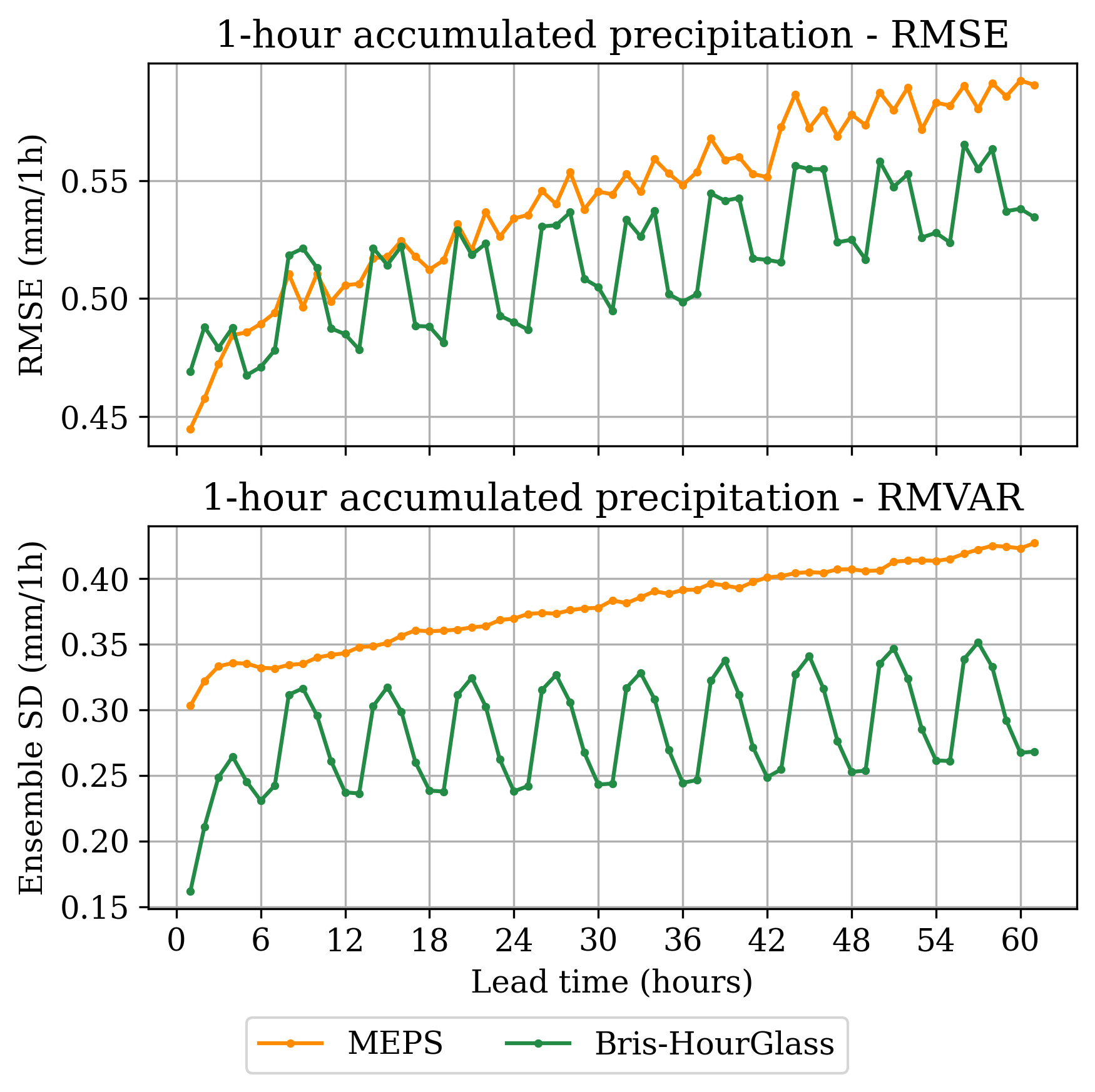}
    \caption {Single-member RMSE and standard deviation (unbiased root mean variance) of 1-hour accumulated precipitation for the MEPS ensemble forecast and Bris-HourGlass. Verification is performed against SYNOP for 2025.}
    \label{fig:precip_rmse-spread}
\end{figure}

For hourly precipitation, the fair CRPS is overall comparable to that of MEPS, but the scores deteriorate near the 6-hourly boundaries. Figure~\ref{fig:precip_rmse-spread} examines the single-member RMSE and the standard deviation of the two ensembles and shows that Bris-HourGlass has lower RMSE near the 6-hourly anchor times. This suggests that the improved fair CRPS toward the middle of the downscaling window is mainly driven by increased ensemble spread there, rather than by uniformly lower errors. Although the model appears to use information from other variables near the boundaries to improve the deterministic skill of hourly precipitation, it also seems to become overconfident in doing so, thus degrading the fair CRPS.

\begin{figure}
    \centering
    \includegraphics[width=\linewidth]{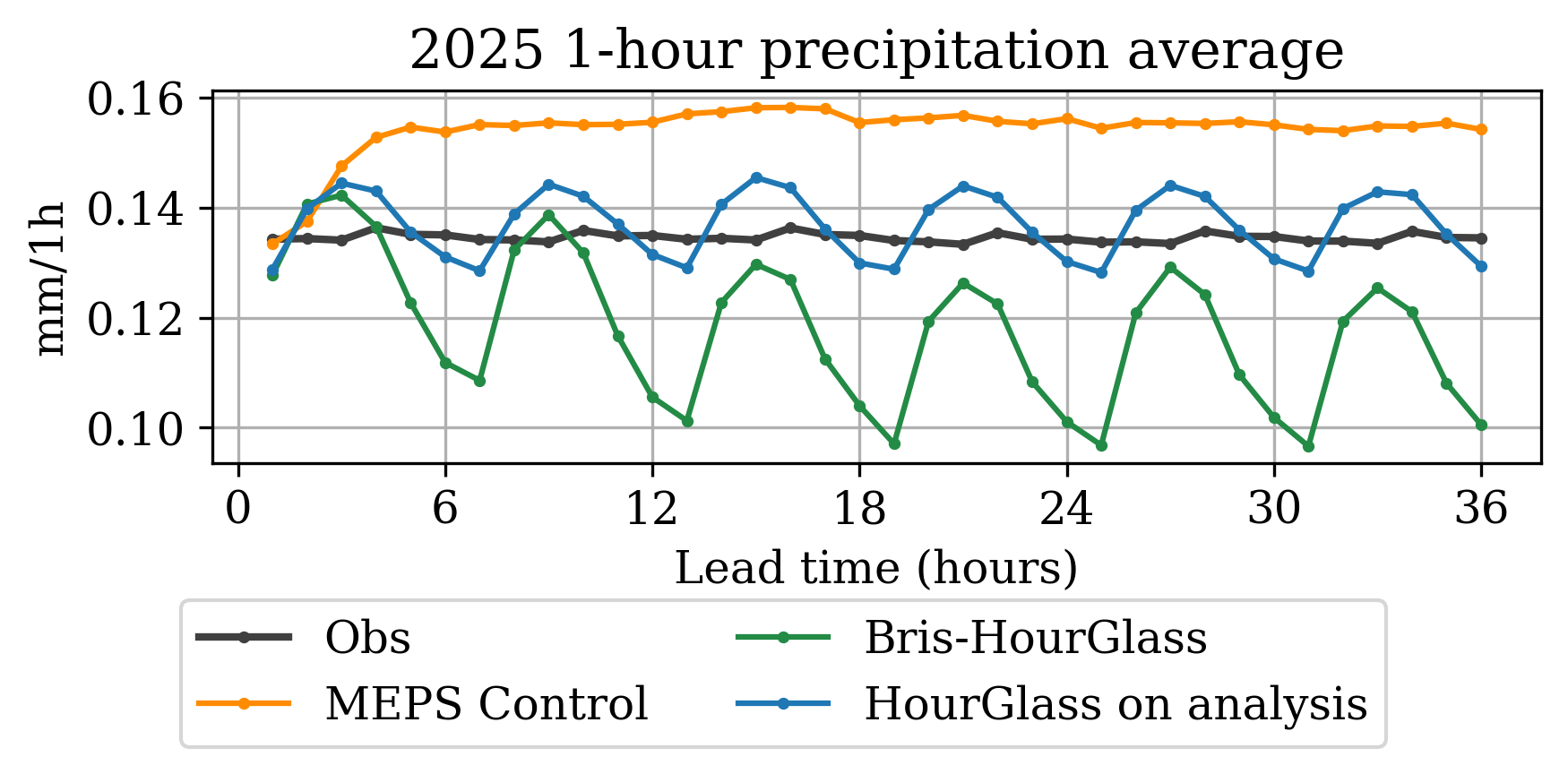}
    \caption{1-hour accumulated precipitation averaged over the 2025 validation period to highlight biases. Compares SYNOP, the MEPS control member, Bris-Hourglass run with the Bris forecaster, and Bris-Hourglass run with the IFS and MEPS analyses at 00, 06, 12 and 18 UTC as temporal boundaries.}
    \label{fig:precip_bias}
\end{figure}

Figure \ref{fig:precip_bias} illustrates how too much dependence on the temporal boundary values for precipitation can create adverse effects on the output. Here, we show the annual average precipitation as a function of lead time when Bris-HourGlass is run with the Bris ensemble forecast and with the MEPS analysis as a temporal boundary. We also include observations and MEPS Control as a reference. Note that precipitation observations are underestimated due to wind-induced gauge undercatch, particularly in winter. The MEPS control forecast starts close to the observations, followed by a spin-up toward higher precipitation values. As Bris-HourGlass is trained on lead times 6-23 of this forecast, the training data consists of these higher precipitation values. Thus, when Bris-HourGlass is run on the MEPS analysis, before this spin-up has occurred, its output tends to increase precipitation toward the middle of the forecast window. Away from the influence of the temporal boundaries, the model relaxes toward precipitation fields that are closer to its learned distribution.

This effect is exacerbated when running Bris-HourGlass with the Bris ensemble forecast as the upstream model. As the initial state at hour 0 is still analysis, precipitation starts out similar to the Bris-HourGlass run on MEPS for the first few hours. However, nearing hour 6, we see a clear reduction in average precipitation, likely coming from a dry bias in the forecasting model. Throughout the forecast, Bris-Hourglass tends more towards its internal learned distribution near the middle of the time window, but it does not fully correct the biased input state.

\begin{figure}
    \centering
    \includegraphics[width=\linewidth]{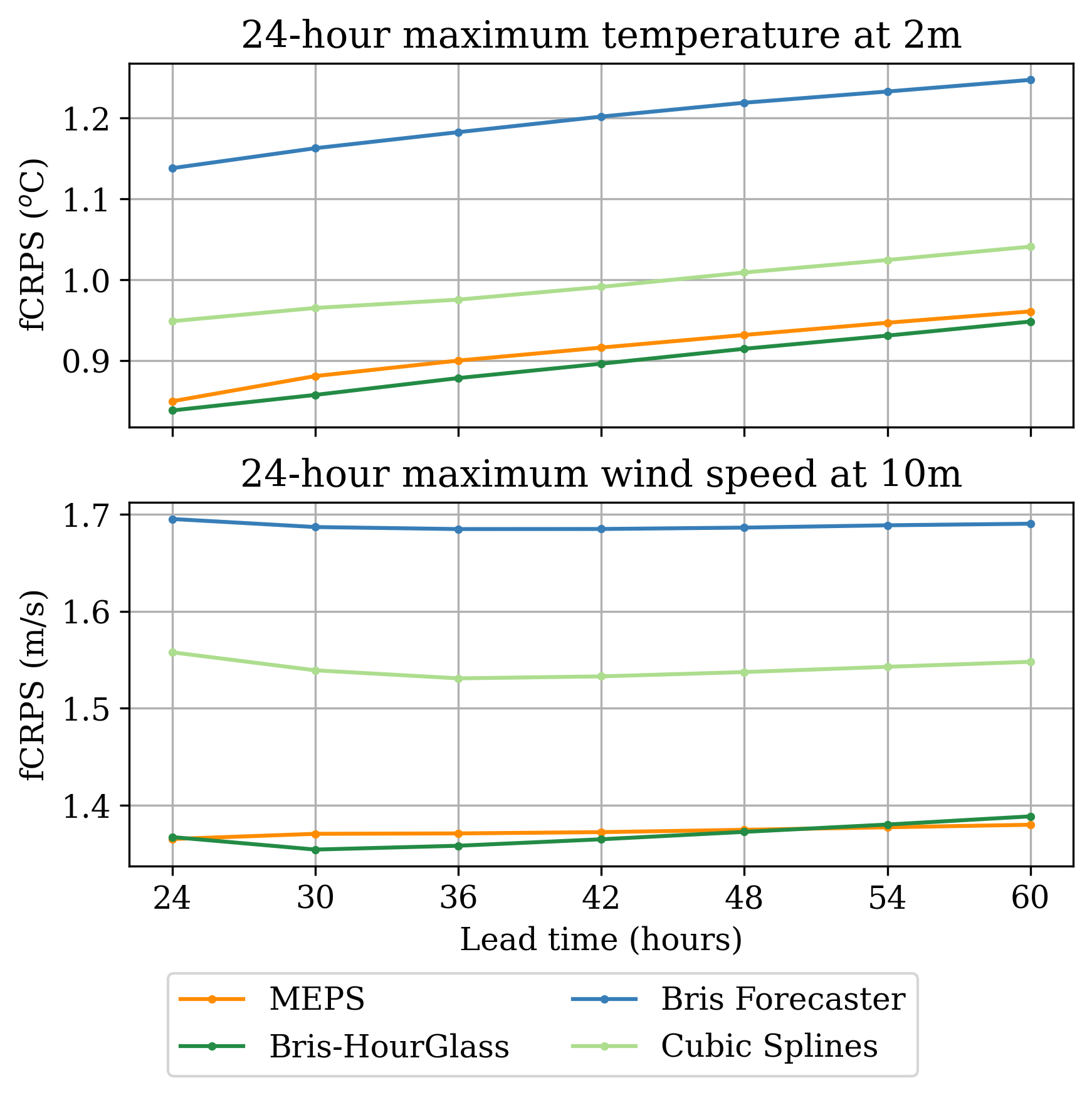}
    \caption{Fair CRPS for 24-hour maximum of 2m temperature and 10m wind speed. Compares the hourly Bris-HourGlass with its 6-hourly upstream Bris forecast and an hourly cubic splines baseline on that forecast. The score of the MEPS ensemble forecast is also included. Verification is performed against SYNOP stations for 2025.}
    \label{fig:24h_accum}
\end{figure}

Hourly forecasts are also necessary for providing good estimates of 24-hour maximum and minimum temperatures and wind speeds, since the actual 24-hour extrema frequently occur between 6-hourly forecast outputs. Figure \ref{fig:24h_accum} shows that Bris-HourGlass is competitive with the MEPS ensemble forecast at capturing the 24-hour maxima, indicating that it is able to model variability within a 6-hour period. In comparison, the cubic splines method improves somewhat on the 6-hourly forecast, but does not succeed in fully capturing the 24-hour peaks in wind speed and temperature. This shows that HourGlass models add value for both short-range forecasts and long-range forecasts.

\section{Case Studies}
\label{sec:case_study}
\subsection{Extra-tropical Storm Amy}

Storm Amy was a severe extratropical cyclone that developed rapidly, southwest of the North Atlantic storm track during the autumn of 2025. By 3rd October, the storm had crossed northern Scotland as a deepening cyclone, before stalling north of Shetland and turning southeast towards Norway. Severe weather warnings for wind and rain were issued across northern Europe for this storm, including red warnings by both Met Eireann and MET Norway. Due to its impact, rapid temporal evolution and strong spatial variability, Storm Amy is a relevant case study for evaluating both Bris-HourGlass and AIFS-HourGlass' ability to reconstruct the timing and intensity of precipitation and wind peaks. 

We first look at Bris-HourGlass and evaluate the model using both meteograms of precipitation and wind speed at selected locations and static maps of the event to assess both the spatial and temporal coherence and propagation of precipitation structures and wind-speed maxima during the storm.

\begin{figure}
    \centering    \includegraphics[width=\linewidth]{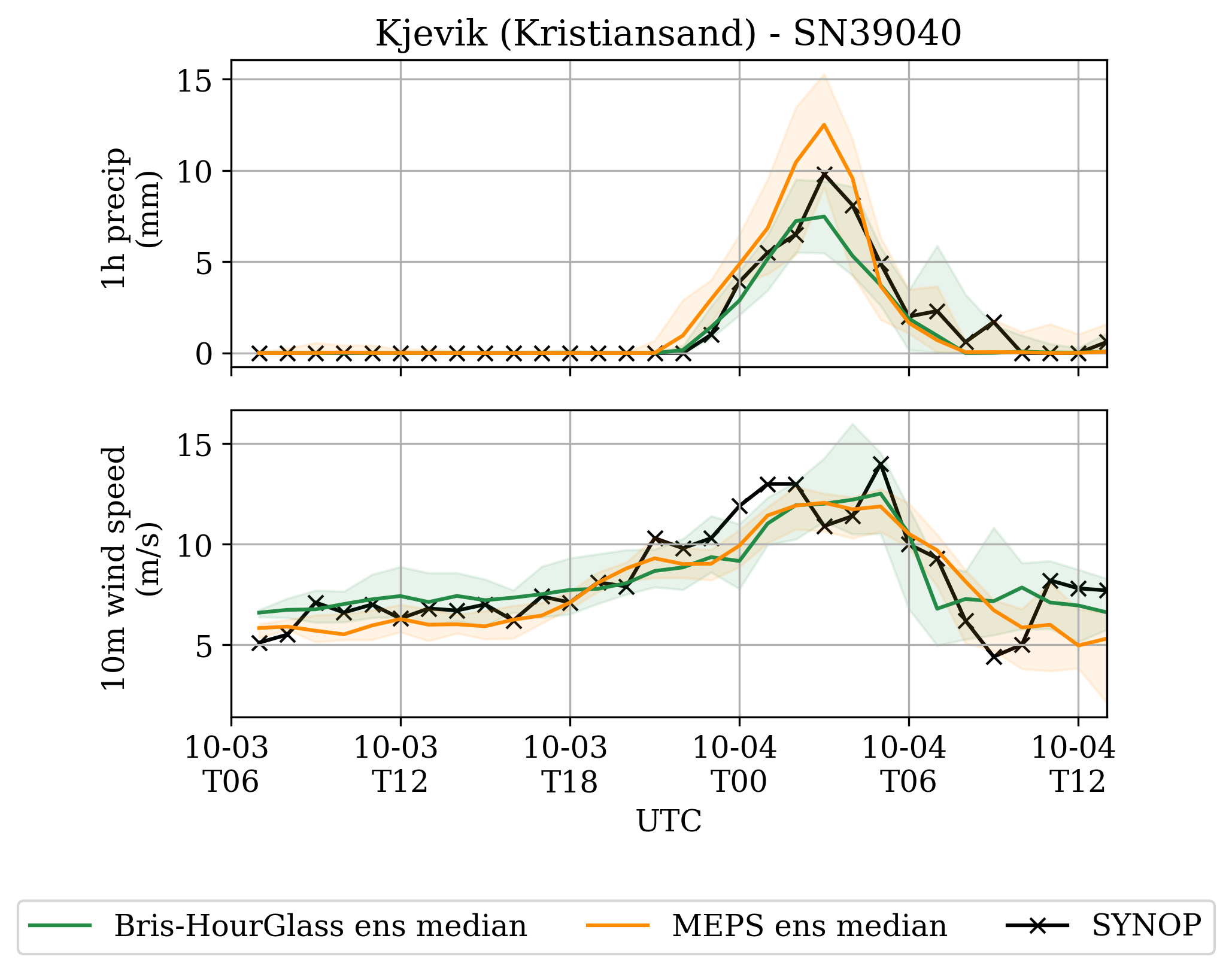}
    \caption{Meteogram for Kjevik airport (SN39040), Norway, during storm Amy. The upper panel shows hourly accumulated precipitation and the lower panel shows 10 metre wind speed. The ensemble median from Bris-HourGlass is compared with the ensemble median of MEPS and SYNOP station observations. Shaded envelopes indicate the 10-90\% quantile ranges. Initialisation time is 2025-10-03 06:00 UTC.}
   \label{fig:amy_meteograms}
\end{figure}

Figure~\ref{fig:amy_meteograms} shows the meteogram at Kjevik airport on the southern coast of Norway during Storm Amy. The forecasts shown are initialised 20 hours before the event, and the station was chosen because the large-scale precipitation system swept over the area between T00 and T06, i.e., within one of the downscaling windows. Both Bris-HourGlass and MEPS capture the timing of onset of precipitation and maximum intensity well. Bris-Hourglass underestimates the peak intensity, which aligns with the precipitation bias discussed in more detail in Sec.~\ref{sec:bris-hourglass}, while MEPS overestimates it. Broadly, the observations fall within the 90\% interval of both models. 
\begin{figure*}
    \centering
    \begin{subfigure}{0.98\linewidth}
        \centering
        \includegraphics[width=\textwidth]{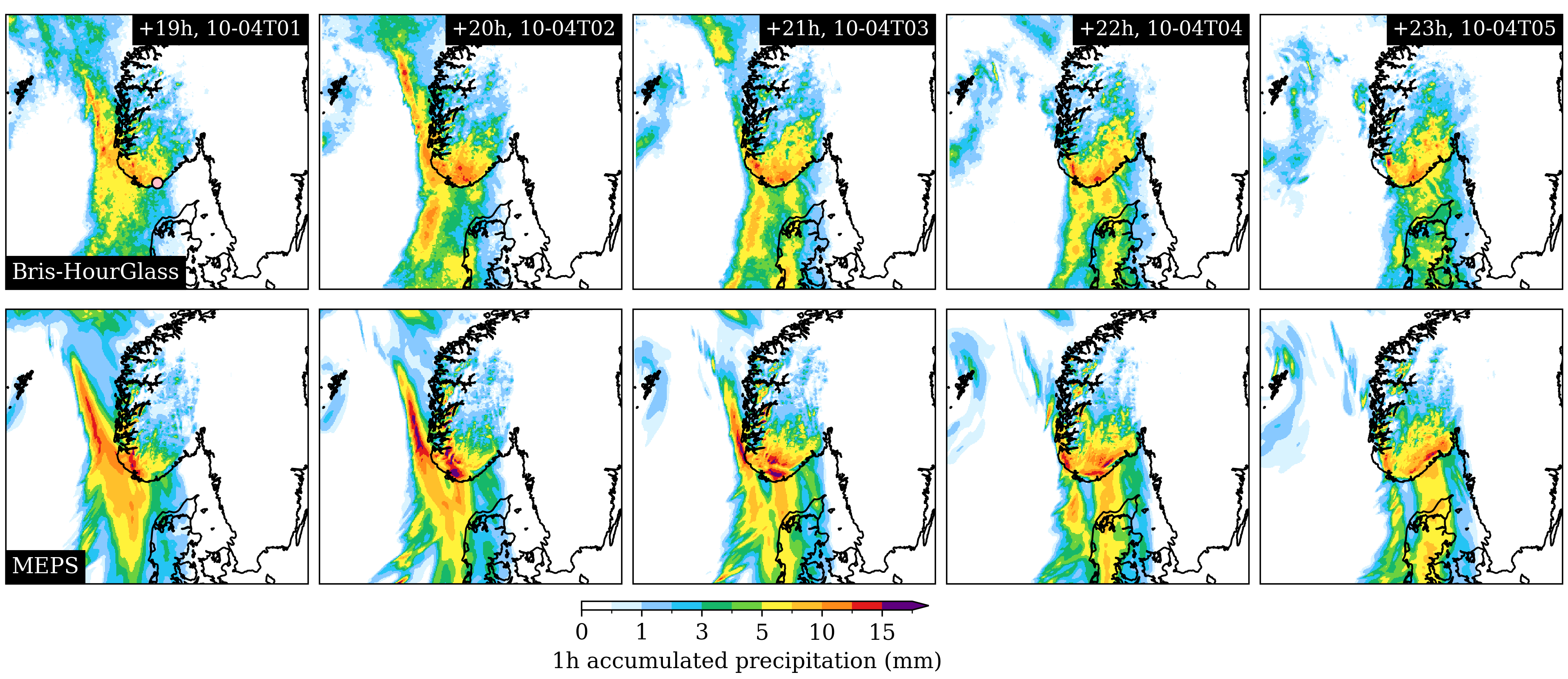}
        \caption{}
        \label{fig:amy_bris_map_a}
    \end{subfigure}

    \begin{subfigure}{0.98\linewidth}
        \centering
        \includegraphics[width=\textwidth]{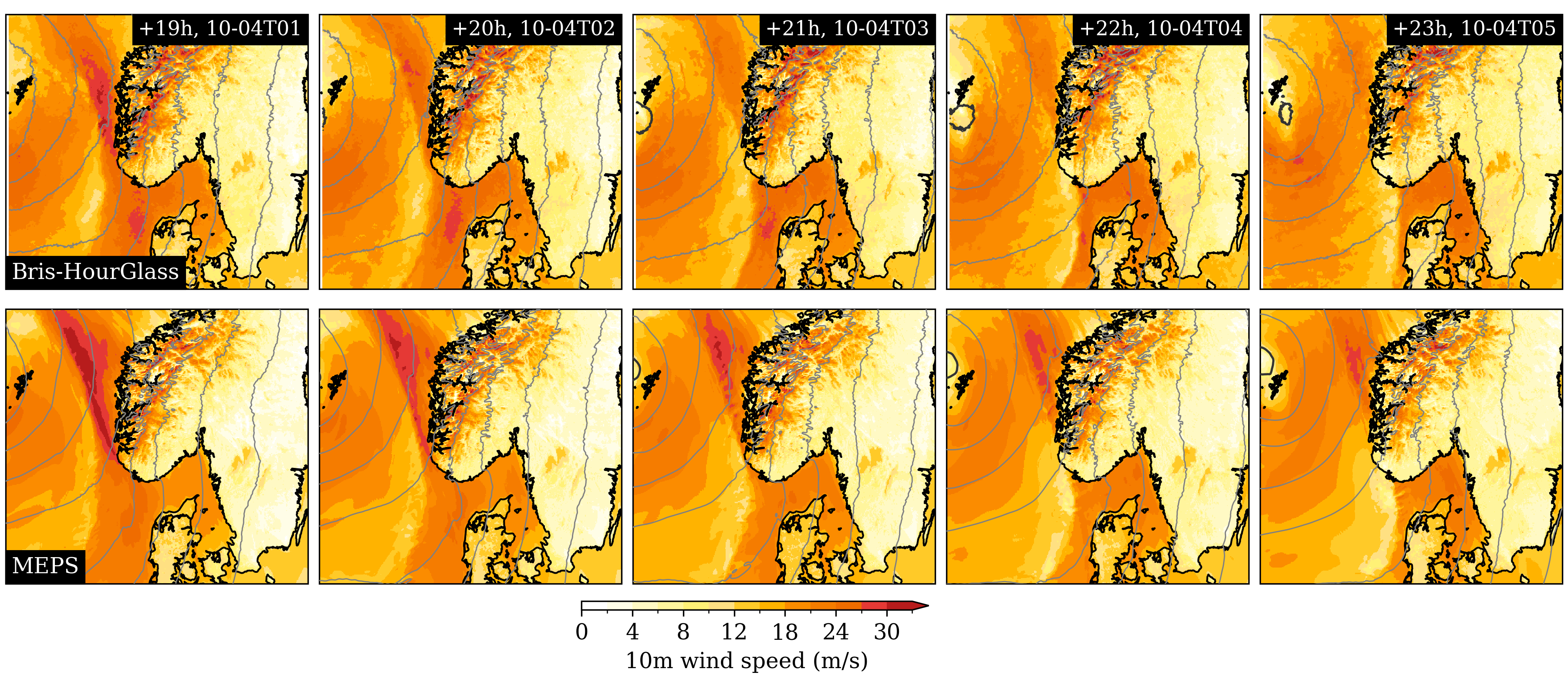}
        \caption{}
        \label{fig:amy_bris_map_b}
    \end{subfigure}
    \caption{Spatio-temporal evolution of (a) hourly precipitation rate and (b) 10 m wind speed and mean sea level pressure (grey contours at 8hPa intervals, with the central low pressure contour at 948hPa highlighted), on 4 October 2025 shown every hour between 01 and 05 UTC. The upper row shows a single member from Bris-HourGlass and the bottom row shows a single member from MEPS. Both forecasts are initialised on 3 October 2025 at 06 UTC. The pink dot in the top left panel in (a) is the location of the Kjevik station, used in Figure \ref{fig:amy_meteograms}.}
   \label{fig:amy_bris_map}
\end{figure*}

Predictions from a single member of Bris-HourGlass and MEPS forecasts during Storm Amy’s peak intensity are shown for precipitation in Figure~\ref{fig:amy_bris_map_a} and 10m wind speed and mean sea level pressure in Figure~\ref{fig:amy_bris_map_b}. For wind speed, Bris-HourGlass preserves the main spatial gradients and the temporal displacement of the high-wind regions, suggesting that the model learns a physically coherent evolution rather than producing independent intermediate fields. For precipitation, the model captures the broad location and movement of the main precipitation system, but the most intense regions have lower amplitudes compared to the forecast from MEPS. This is consistent with the meteogram in Figure~\ref{fig:amy_meteograms} and discussion in Sec.~\ref{sec:bris-hourglass}.

\begin{figure*}
    \centering
    \begin{subfigure}{0.98\linewidth}
        \centering
        \includegraphics[width=\textwidth]{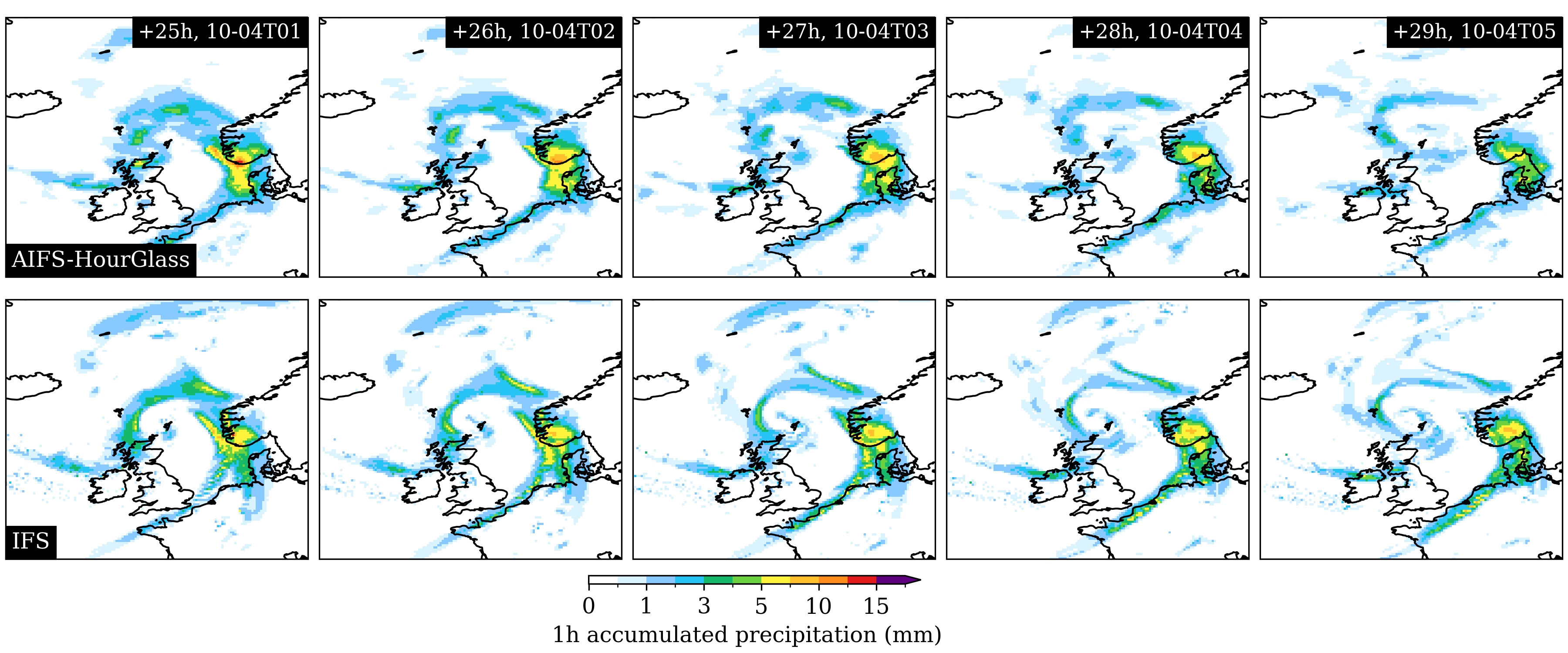}
        \caption{}
        \label{fig:amy_aifs_map_a}
    \end{subfigure}

    \begin{subfigure}{0.98\linewidth}
        \centering
        \includegraphics[width=\textwidth]{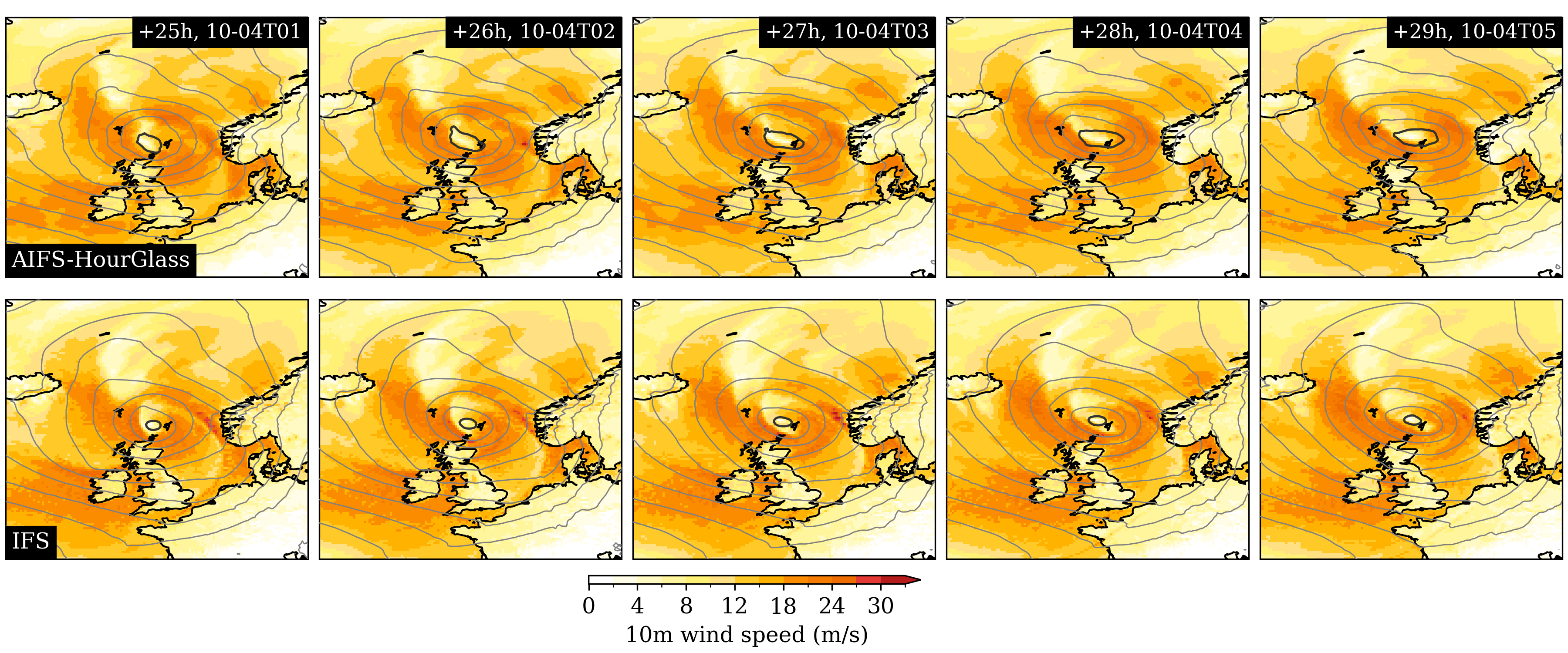}
        \caption{}
        \label{fig:amy_aifs_map_b}
    \end{subfigure}
    \caption{Spatio-temporal evolution of (a) hourly precipitation rate and (b) 10m wind speed and mean sea level pressure (grey contours at 8-hPa intervals, with the central low pressure contour at 948 hPa highlighted), on 4 October 2025 shown every hour between 01 and 05 UTC. The upper row shows a single member from AIFS-HourGlass and the bottom row shows a single member from IFS. Both forecasts are initialised on 3 October 2025 at 00 UTC. All fields are conservatively interpolated onto a common $0.25^\circ\times 0.25^\circ$ grid for comparability.}
   \label{fig:amy_aifs_map}
\end{figure*}

Since Storm Amy affected large parts of northwestern Europe beyond the Bris domain, we additionally evaluate the performance of the AIFS-HourGlass forecasts for the same variables considered previously. Figure~\ref{fig:amy_aifs_map_a} shows the sub-six-hourly evolution of precipitation forecasts from one ensemble member of AIFS-HourGlass and IFS. Both models produce similar overall precipitation patterns, with rainbands closely aligned with the storm's frontal structures. In contrast to Bris-HourGlass, the maximum precipitation values for AIFS-HourGlass are comparable to those in the IFS forecast. A notable difference with IFS, however, is that AIFS-HourGlass tends to generate broader rainbands. Given comparable precipitation maxima, this results in smoother spatial gradients than in IFS. A comparison of the surface wind forecasts in Figure~\ref{fig:amy_aifs_map_b} reveals high similarity in the synoptic-scale cyclone structure. However, AIFS-HourGlass underestimates the strong wind gradients near the cyclone centre and fails to capture the wind speed maxima predicted by IFS west of Norway. As these differences persist across all sub-six-hourly lead times shown, they are likely attributable to deficiencies in the six-hourly AIFS model (in particular its coarser spatial resolution relative to IFS) rather than to the temporal downscaling. Moreover, the physically consistent relationship between the pressure-gradient structure and surface wind speeds suggests that AIFS-HourGlass generates dynamically coherent intermediate states at the synoptic scale, rather than merely generating smooth transitions between the coarse forecast anchors.

Overall, the case study indicates that the model can generate temporally consistent intermediate states during a severe storm, both at local station scale and in the larger-scale spatial evolution. The results also highlight the remaining challenge of representing sharp precipitation features and local wind-speed extremes.

\subsection{Convective episode in the United States Southern Plains}
\label{sec:convective_case}
\begin{figure}[tbp]
    \centering

    \begin{subfigure}[t]{0.98\linewidth}
        \centering
        \includegraphics[width=\linewidth]{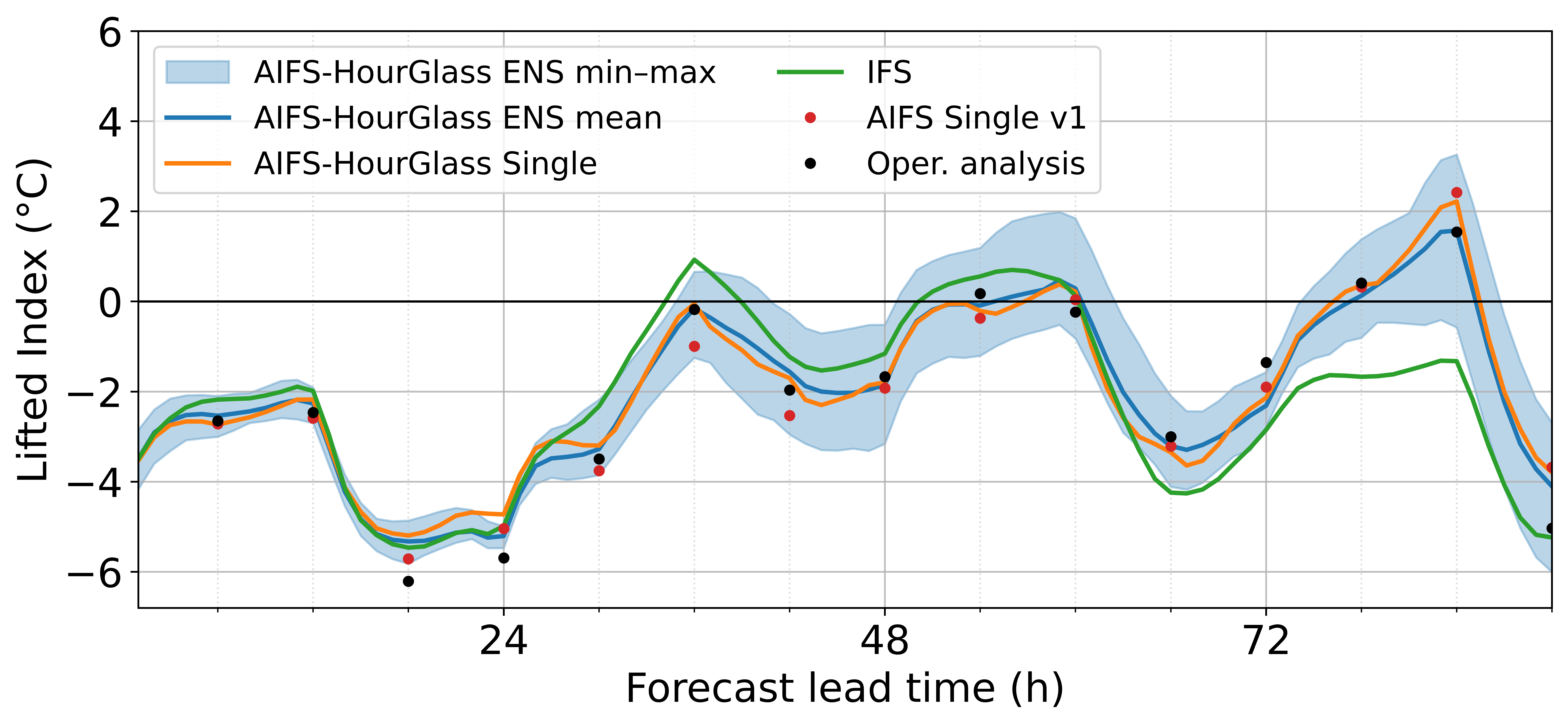}
        \caption{}
        \label{fig:convective_case_LI_evol}
    \end{subfigure}
    \vspace{0.5em}
    \begin{subfigure}[t]{0.98\linewidth}
        \centering
        \includegraphics[width=\linewidth]{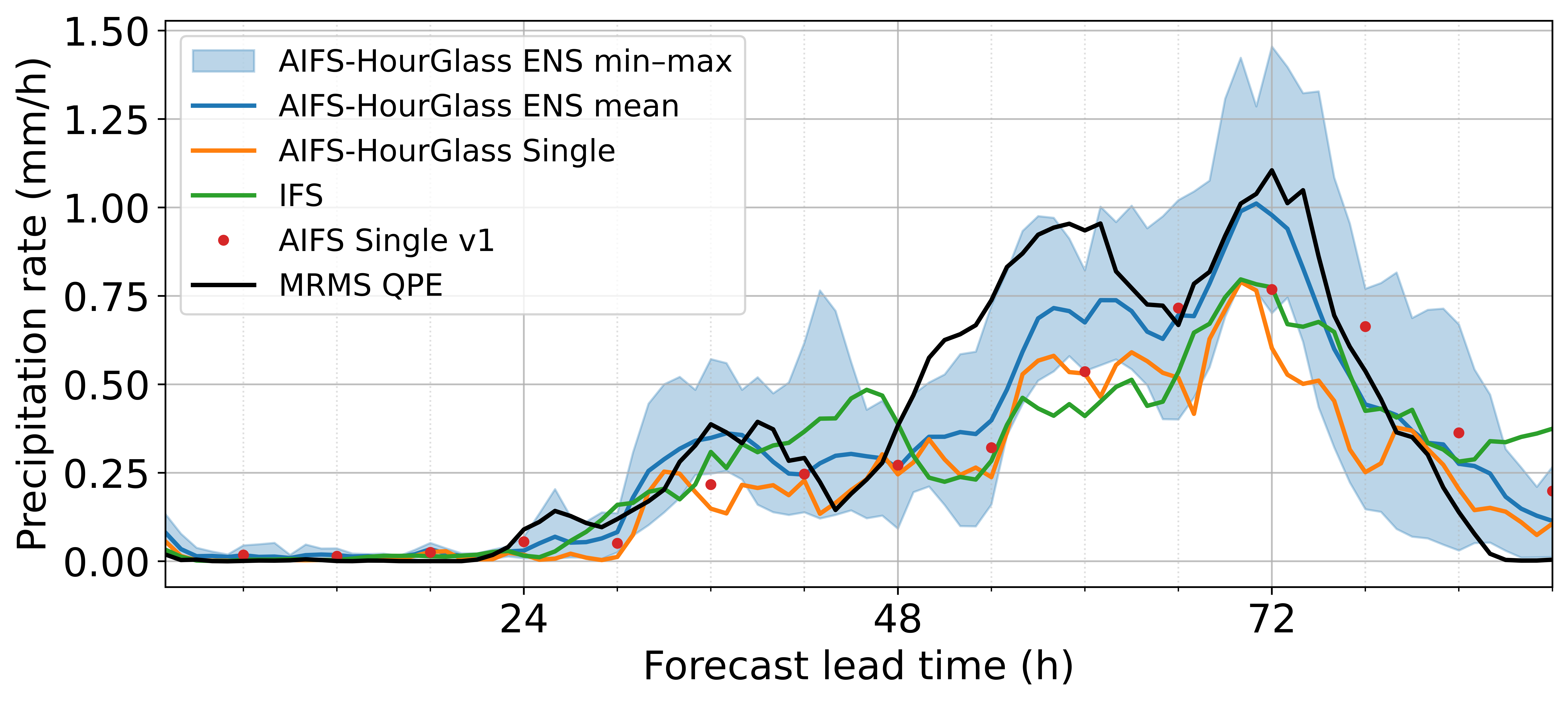}
        \caption{}
        \label{fig:convective_case_tp_evol_area-mean}
    \end{subfigure}
    \vspace{0.5em}
    \begin{subfigure}[t]{0.98\linewidth}
        \centering
        \includegraphics[width=\linewidth]{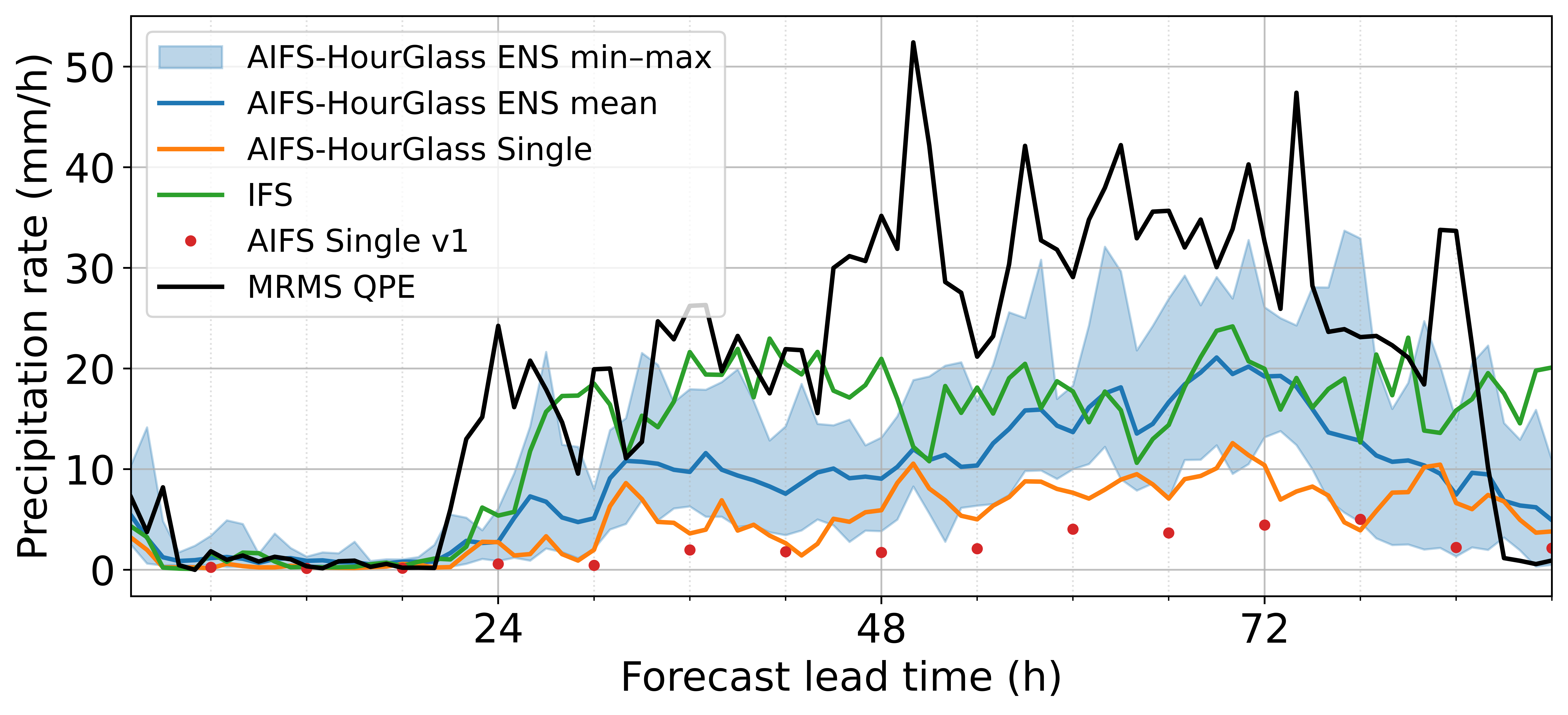}
        \caption{}
        \label{fig:convective_case_tp_evol_area-max}
    \end{subfigure}

    \caption{Time series of precipitation and atmospheric instability over $30$--$40^\circ$N, $258$--$270^\circ$E from forecasts initialised at 00 UTC 28 April 2025. (a) Area-mean Lifted Index (LI, $^\circ$C), with negative values indicating increased atmospheric instability. (b) Area-mean and (c) area-maximum hourly total precipitation rate (mm h$^{-1}$). Shaded regions indicate ensemble spread (min--max), while solid lines denote ensemble mean and deterministic forecasts.}
    \label{fig:convective_case_time_evol}
\end{figure}

\begin{figure*}
    \centering
    \includegraphics[width=\textwidth]{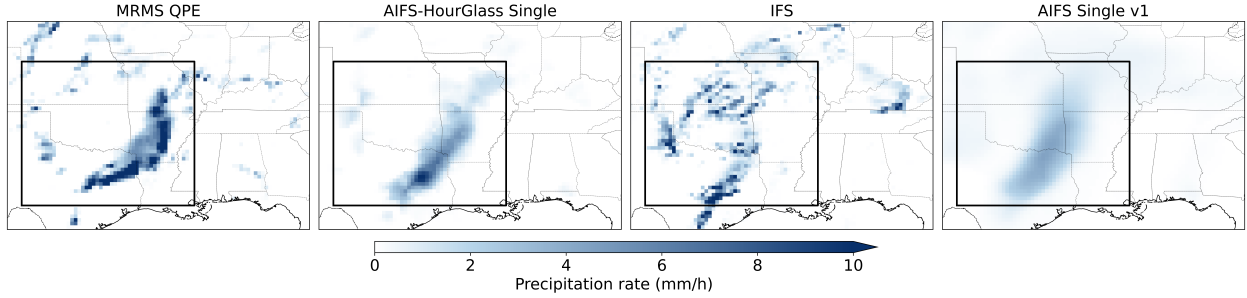}
    \caption{Spatial distribution of hourly total precipitation rate (in mm/h) from observations (MRMS QPE), and AIFS-HourGlass Single, IFS, and AIFS Single forecasts at 72~h lead time, valid at 00 UTC 1 May 2025. All fields are conservatively interpolated onto a common $0.25^\circ\times 0.25^\circ$ grid for comparability, and the precipitation rate of AIFS Single is divided by 6 to match the hourly interval. The black rectangle delineates the domain ($30$--$40^\circ$N, $258$--$270^\circ$) used for the time series of the spatial statistics presented in Fig.~\ref{fig:convective_case_time_evol}.}
   \label{fig:convective_tp_maps}
\end{figure*}

In the last days of April 2025, the Southern Plains of the United States were affected by a series of convective systems. Thunderstorm activity first initiated over New Mexico and northwestern Texas around 21 UTC on 28 April (late afternoon local time), with new convective cells repeatedly developing over the same region during the following days. As convection propagated northeastward, some of the cells organised into larger mesoscale convective systems (MCSs). The most intense phase occurred on 30 April, when a long-lived system formed at around 07 UTC (predawn local time) and subsequently evolved into an eastward-moving line of severe convection. This episode of extreme convective activity produced heavy rainfall and flash flooding.

The pronounced spatial gradients, localised precipitation maxima, and rapid temporal evolution characteristic of such MCSs provide a demanding test case for assessing whether temporal downscaling to hourly resolution captures observed sub-six-hour variability under strongly convective conditions. We evaluate AIFS-HourGlass forecasts initialised three days prior (28 April 2025 00UTC) to the peak of the convective episode. The verification domain spans $30$--$40^\circ$N and $258$--$270^\circ$E, encompassing the region where the convection was initiated, organised and ultimately produced its highest impact.  Verification of hourly forecasts is performed using the ECMWF operational analysis and the Multi-Radar Multi-Sensor Quantitative Precipitation Estimation (MRMS QPE) \citep{zhang2016mrms}.

To assess the potential for convection, the Lifted Index \citep[LI;][]{galway1956LI} is computed as a single-level stability measure. It compares the environmental temperature at 500 hPa with that of an air parcel dry-adiabatically lifted to its lifted condensation level and then moist-adiabatically to 500 hPa, resulting in a single indicator of convective instability. More negative LI values indicate greater convective instability and a higher potential for deep convection. The observed area-averaged LI shown in Figure~\ref{fig:convective_case_LI_evol} reveals that convectively favourable conditions on the larger mesoscales consistently emerge at the onset of the second half of the day. Instability then increases markedly over the following 4-8 h before slightly stabilising again during the final hours of the day. The phasing of the extrema in this diurnal cycle is generally well predicted by all models, and the magnitude remains comparable across models and with the analysis. Moreover, periods such as 12-18 h and 24-30 h demonstrate that AIFS-HourGlass does not merely interpolate between the 6‑h intervals, but captures nonlinear transitions similar to the IFS, enabling peaks to be resolved at sub‑six‑hour time steps (e.g. just after 66 h). The strong agreement with the operational analysis confirms that the AIFS-HourGlass can represent the bulk thermodynamic conditions necessary for convective activity organised up to the large mesoscale in a physically consistent manner.

The time series of the area-averaged precipitation rate during the convective episode is shown in Figure \ref{fig:convective_case_tp_evol_area-mean}. Apart from brief periods of slight weakening, the observed precipitation generally increases from late 28 April to the end of 30 April. This overall evolution is captured by both AIFS-HourGlass and IFS forecasts. However, all models underestimate the area‑averaged precipitation rates, particularly during forecast day 3 (30 April). Among the models considered, the AIFS-HourGlass ensemble most closely predicts the observed evolution. The ensemble mean agrees best with the observations from approximately 48 h lead time through the peak phase at the end of 30 April and into the morning of 1 May. In contrast, the AIFS-HourGlass Single and the IFS remain near the minimum of the ensemble range during the most active period, with the AIFS-HourGlass Single even yielding lower mean precipitation rates than the six‑hourly AIFS Single during that time, consistent with Figure \ref{fig:tp_dist}.

The area-maximum time series presented in Figure~\ref{fig:convective_case_tp_evol_area-max} exhibits markedly stronger and higher-frequency fluctuations, reflecting the challenge of predicting the most extreme values among many rapidly evolving convective systems. The observations contain short-lived local maxima exceeding 40--50 mm h$^{-1}$, mainly on 30 April, whereas all models substantially underestimate these extremes. Even the most intense members of the AIFS-HourGlass ENS only occasionally reach 30--35 mm h$^{-1}$. The IFS maximum remains comparatively steady throughout the episode, fluctuating around values slightly below 20 mm h$^{-1}$. In contrast, the AIFS-HourGlass ENS mean follows the observed increase from 24 to 72 h, although the area-maximum precipitation is clearly underestimated. The AIFS-HourGlass Single improves upon the six-hourly AIFS Single by generally shifting the forecast towards higher precipitation maxima, but still performs worse than most members of the AIFS-HourGlass ensemble, likely because of the smooth AIFS Single fields used in the inputs for AIFS-HourGlass Single. Furthermore, the temporal downscaler appears to push sub-six-hourly values towards higher intensities in a physically consistent manner, although the extent of this enhancement remains constrained by the six-hourly input fields.

Figure~\ref{fig:convective_tp_maps} compares precipitation rate maps at the peak convective activity (72 h lead time) with observations. At this time, a southwest-northeast-oriented, back-bent line of cells is observed, generating strong localised precipitation while propagating eastward. Both the hourly and six-hourly AIFS Single forecasts produce a correctly oriented broad band of enhanced precipitation, with the AIFS-HourGlass Single exhibiting more localised and higher extremes than the smoother six-hourly AIFS Single. However, both AIFS Single versions fail to model the back-bent deformation, which likely arises from processes occurring at smaller scales than those resolved. While the IFS deviates more from the observed position and overall shape of the precipitation band, it generates more localised and intense precipitation cells. These cells are, however, more fragmented and spread across the verification domain, indicating weaker convective organisation.

The case study shows that temporal downscaling adds meaningful sub-six-hourly information to the AIFS forecasts under strongly convective conditions on the larger mesoscale. It captures the broad evolution of convective episodes, including diurnal and sub-diurnal variability in both bulk thermodynamic conditions and mean precipitation rates. In addition, it produces precipitation structures that are closer to observations than the smoother AIFS Single and the more fragmented and less organised IFS fields. However, the largest local precipitation extremes remain underestimated, particularly in the area-maximum diagnostic, in part due to the coarser spatial resolution of the AIFS models, but also due to the smooth fields used in input for AIFS-HourGlass Single.

\section{Conclusion}\label{sec:conclusion}

In this work, we introduce HourGlass, a probabilistic methodology for reconstructing the temporal evolution between two states provided by a forecasting model. This temporal downscaling methodology is demonstrated in both the global (AIFS-HourGlass) and regional (Bris-HourGlass) settings. Our results show that in both cases, HourGlass provides hourly forecasts that largely preserve the skill of the underlying forecasting systems.

A key contribution of this work is the use of probabilistic training based on an almost-fair CRPS objective combined with temporal consistency constraints. As a result, HourGlass generates coherent hourly trajectories that preserve realistic spatial variability throughout the downscaling window, rather than the temporally smoothed intermediate states seen when training deterministically. Equally important, we have shown that training HourGlass models on forecast datasets provides a practical way to learn long consistent temporal trajectories. These provide an alternative when consistent hourly analysis/reanalysis datasets are unavailable, while still allowing the temporal downscaler to benefit from the skill of analysis-trained data-driven forecasting systems.

Verification against SYNOP stations shows that both the HourGlass models preserve the skill of the upstream forecasting models and closely follow the temporal behaviour of the hourly NWP benchmarks. The case studies further demonstrate that both models can reproduce physically consistent temporal evolution during rapidly evolving weather events, including extratropical cyclones and organised convection. They capture the broad timing, displacement, and spatial organisation of these events. HourGlass successfully improves the spatial realism of precipitation fields over the upstream forecasts

Despite these encouraging results, like many data-driven forecasting models, the HourGlass models still underestimate sharp local precipitation maxima. They remain limited by both the distribution of the forecasts on which they are trained and the quality of the forecasts at the 6-hourly boundaries. Thus, improvements to the upstream forecasts are expected to translate directly into improved hourly forecasts. 

To improve the HourGlass models, future work will investigate more robust probabilistic training, particularly near the temporal boundaries. Additionally, improvements to the HourGlass loss function will be investigated, such as introducing a multi-scale loss and adding terms to discourage insufficient spread near the boundaries.

Overall, the skilful temporally coherent hourly forecasts produced by HourGlass demonstrate its potential to bridge the gap between coarse-temporal-resolution data-driven forecasts and the hourly products required in operational regional and global forecasting.

\subsection*{Code availabilty}
The code and training configurations used to train the HourGlass models in this work are available at \url{https://github.com/ecmwf/anemoi-core/tree/feature/ens_interp}.
\subsection*{Acknowledgments}
The authors gratefully acknowledge all contributors to the Anemoi framework, in particular Dieter Van den Bleeken for his work on the multi-output set-up. 

The authors further gratefully acknowledge the EuroHPC Joint Undertaking for awarding the project EHPC-REG-2024R02-079 access to the EuroHPC supercomputer LEONARDO, hosted by CINECA (Italy) and the LEONARDO consortium through an EuroHPC Regular Access call. The authors also gratefully acknowledge the Gauss Centre for Supercomputing e.V. (www.gauss-centre.eu) for funding this project by providing computing time on the GCS Supercomputer JUPITER at Jülich Supercomputing Centre (JSC). 

Part of the work presented in this paper has been produced in the context of the European Union’s Destination Earth Initiative and relates to tasks entrusted by the European Union to the European Centre for Medium-Range Weather Forecasts implementing part of this Initiative with funding by the European Union. Views and opinions expressed are those of the author(s) only and do not necessarily reflect those of the European Union or the European Commission. Neither the European Union nor the European Commission can be held responsible for them.

Part of the work is supported by the ECMWF Machine Learning Pilot Project, a European collaborative initiative to advance data-driven weather prediction.

\newpage
\bibliographystyle{plainnat}
\bibliography{references}  



\end{document}